\documentclass[twocolumn,tighten]{aastex62}
\RequirePackage{rotating}
\usepackage{graphicx}
\usepackage{grffile}
\usepackage{subfigure}
\usepackage{amsmath}

\usepackage[normalem]{ulem}

\graphicspath{{./}{figures/}}

\usepackage{soul} % for crossing out text

\submitjournal{ApJ}

\shorttitle{Ion drift in the solar atmosphere}
\shortauthors{Mart\'inez-Sykora et al.}

\usepackage{natbib}
\begin{document}

\title{On the velocity drift between ions in the solar atmosphere}

\correspondingauthor{Juan Mart\'inez-Sykora}
\email{juanms@lmsal.com}

\author{Juan Mart\'inez-Sykora}
\affil{Lockheed Martin Solar \& Astrophysics Laboratory,
3251 Hanover St, Palo Alto, CA 94304, USA}
\affil{Bay Area Environmental Research Institute,
NASA Research Park, Moffett Field, CA 94035, USA}
\affil{Rosseland Centre for Solar Physics, University of Oslo,
P.O. Box 1029 Blindern, NO0315 Oslo, Norway}

\author{Mikolaj Szydlarski}
\affil{Rosseland Centre for Solar Physics, University of Oslo,
P.O. Box 1029 Blindern, NO0315 Oslo, Norway}
\affil{Institute of Theoretical Astrophysics, University of Oslo, P.O. Box 1029 Blindern, N-0315 Oslo, Norway}

\author{Viggo H. Hansteen}
\affil{Lockheed Martin Solar \& Astrophysics Laboratory,
3251 Hanover St, Palo Alto, CA 94304, USA}
\affil{Bay Area Environmental Research Institute,
NASA Research Park, Moffett Field, CA 94035, USA}
\affil{Rosseland Centre for Solar Physics, University of Oslo,
P.O. Box 1029 Blindern, NO0315 Oslo, Norway}
\affil{Institute of Theoretical Astrophysics, University of Oslo, P.O. Box 1029 Blindern, N-0315 Oslo, Norway}

\author{Bart De Pontieu}
\affil{Lockheed Martin Solar \& Astrophysics Laboratory,
3251 Hanover St, Palo Alto, CA 94304, USA}
\affil{Rosseland Centre for Solar Physics, University of Oslo,
P.O. Box 1029 Blindern, NO0315 Oslo, Norway}
\affil{Institute of Theoretical Astrophysics, University of Oslo, P.O. Box 1029 Blindern, N-0315 Oslo, Norway}

\newcommand{\eg}{{\it e.g.,}} 
\newcommand{\ie}{{\it i.e.,}} 
\newcommand{\myemail}{juanms@lmsal.com}
\newcommand{\komment}[1]{\texttt{#1}}
\newcommand{\pref}{\protect\ref}
\newcommand{\soho}{{\em SOHO{}}}
\newcommand{\sdo}{{\em SDO{}}}
\newcommand{\stereo}{{\em STEREO{}}}
\newcommand{\iris}{{\em IRIS{}}}
\newcommand{\hinode}{{\em Hinode{}}}
\newcommand{\ebysus}{{\em Ebysus{}}}
\newcommand{\bifrost}{{\em Bifrost{}}}
\newcommand{\wc}{multi-ion cyclotron}
\newcommand{\hp}{H$^{+}$}
\newcommand{\hep}{He$^{+}$}
\newcommand{\cp}{C$^{+}$}
\newcommand{\hepp}{He$^{++}$}
\newcommand{\jms}[1]{\color{black}{#1}}
\newcommand{\vhh}[1]{\color{blue}{#1}}
\newcommand{\ms}[1]{\color{magenta}{#1}}
\newcommand\msrm{\bgroup\markoverwith{\textcolor{magenta}{\rule[0.5ex]{2pt}{0.4pt}}}\ULon}

\begin{abstract}

The solar atmosphere is composed of many species which are populated at different ionization and excitation levels. The upper chromosphere, transition region and corona are nearly collisionless. Consequently, slippage between, for instance, ions and neutral particles, or interactions between separate species, may play important roles. We have developed a 3D multi-fluid and multi-species numerical code (\ebysus ) to investigate such effects. \ebysus\ is capable of treating species (\eg\ hydrogen, helium etc) and fluids (neutrals, excited and ionized elements) separately, including non-equilibrium ionization, momentum exchange, radiation, thermal conduction, and other complex processes in the solar atmosphere. Treating different species as different fluids leads to drifts between different ions and an electric field that couples these motions. The coupling for two ionized fluids can lead to an anti-phase rotational motion between them. Different ionized species and momentum exchange can dissipate this velocity drift, \ie\ convert wave kinetic energy into thermal energy.
High frequency Alfv\'en waves {\jms driven by, e.g., reconnection} thought to occur in the solar atmosphere, can drive such multi-ion velocity drifts.
\end{abstract}


\section{Introduction}

To study the outer atmosphere of the Sun and in particular, the interaction between the magnetic field and plasma a single fluid magnetohydrodynamic (MHD) treatment is often chosen. {\jms This is usually sufficient, however, single-fluid MHD process timescales may become comparable to collision frequencies or the presence of neutrals may dominate the dissipation processes in the atmosphere.} As one progresses upwards in the solar atmosphere, and especially into the relatively cold, tenuous solar chromosphere, a multi-fluid magnetohydrodynamic (MHD) treatment of the plasma and magnetic field becomes increasingly relevant in order to handle many aspects of the weakly ionized, weakly collisional medium. For instance, the high-fraction of neutral particles in the chromosphere may play an important role, and drift velocities between different fluids may be too important to neglect \citep{Vernazza:1981yq,Fontenla:1990yq,Leake:2014fk,Martinez-Sykora:2015lq,Martinez-Sykora:2017rb,Ballester:2018fj}. Some studies have already gone beyond a single fluid approach, and these can be categorized into two groups: First, in a weakly ionized plasma with ion-neutral collision time-scales small enough, one can assume that ion-neutral interactions effects can be approximated through an expansion of Ohm's law \citep[see][amongst others]{cowling1957,Braginskii:1965ul,Parker:2007lr} by including ambipolar diffusion \citep{Biermann:1950yg} into the single fluid MHD equations, \citep[\eg ][]{Leake:2005rt,Cheung:2012uq,Khomenko:2012bh,Martinez-Sykora:2012uq,Leake:2013dq,Martinez-Sykora:2017sci,Nobreg-Siverio:2020AA...633A..66N}. The second group of studies treats ions and neutrals as two completely different fluids \citep[\eg][]{Brandenburg:1994qy,Brandenburg:1995jb,Lazarian:2004xi,Leake:2013qz,Alvarez-Laguna:2016ty,Alvarez-Laguna:2017rz,Maneva:2017zl} with separate velocities and energies, only coupled through  collisions.

The solar atmosphere consists of large variety of species and most previous studies have typically assumed that species can be treated as a single or two fluids. \citet{Chapman:1970yx} derived the single fluid equations from a multi-species description. This required the assumption that all species move in unison, which it is not necessarily the case in a weakly collisional environment. Furthermore physical processes are different depending on the ionization state of the species under consideration (see Sections~\ref{sec:methods} and~\ref{sec:iondf}). 
\citet{Khomenko:2014nr} and \citet{Ballester:2018fj} describe the full set of MHD equations separately for each species. This description is better suited to describing the plasma when collisional interactions between species are weak. 

Under the conditions where collisional interactions between species are weak, a physical processes due to electromagnetic interactions between ionized species become important. Alfv\'en waves exhibit one mode that does not appear in single-ion plasma \citep[a fluid with a single ionized species][]{Weber:1973ap,Cramer:2001hl}. The frequency of this mode is known as weighted average ion-cyclotron frequency of the different ionized species participating in the wave motion (in this paper, we refer to this as the \wc\ frequency). The mode can be dissipated by the momentum exchange between species \citep{Isenberg:1982xq,Hollweg:2002il,Rahbarnia:2010vt,Martinez-Gomez:2016im,Martinez-Gomez:2017yb}. Ionized species are accelerated by a non-dissipative wave pressure and a dissipative heating occurs \citep{Isenberg:1982xq}. \citet{Li:2007zn,Li:2008qf} investigated  effects due to interactions between ionized species for these non-Wentzel-Kramers-Brillouin (WKB) Alfv\'en waves. Further studies of multi-ion fluids on waves and anisotropic magnetic plasma instabilities have been carried out by \citet{Demars:1979vb,Olsen:1999nb,Dzhalilov:2008sd}. All these studies have been done analytically with simplifications or/and linear wave approximations. 

Aside from solar physics, other fields have investigated multi-ion plasmas. \citet{Schunk:1977rx,Barakat:1982mf} review the research on mixing multi-species in the anisotropic ionosphere and \citet{Ganguli:1996go} in the polar magnetosphere \citep[see also][]{Demars:1994wk}. \citet{Echim:2011rv,Abbo:2016ga} review multi-fluids modeling and observations in the solar wind, and \citet{Krticka:2000om,Krticka:2001kn} in other stellar atmospheres. The latter two studies did not explicitly solve the equations describing magnetic field evolution. Other studies have shown the importance of including the induction equation \citep[\eg][]{Isenberg:1982xq,Martinez-Gomez:2016im} -- the Larmor frequency of two different ionized species depends on the weighted atomic mass of the ionized atoms or molecules participating in the wave motion. \citet{Xie:2004JGRA..109.8103X,Ofman:2005JGRA..110.9102O} investigated wave dispersion in collisionless multi-ion plasmas. For this, they used nonlinear one-dimensional hybrid kinetic simulations of the multi-ion plasma to investigate high-frequency wave-particle interactions in the solar wind. \citet{Koch:2005ao} found that measurements of the resonant frequency of the ion-cyclotron wave can resolve the composition of the matter in the plasma, as the frequency depends on the abundances of the fluids comprising the plasma. They used the ElectroSpray Ionization Fourier Transform Ion Cyclotron Resonance Mass Spectrometry (ESI FT-ICR-MS) to reveal the composition of a dissolved organic matter, which was up to that point unknown. 

In order to investigate the issues briefly presented above, we have developed the multi-fluid and multi-species (MFMS) numerical code: \ebysus. In the next section we briefly describe the physics implemented and numerical methods used in that code. Section~\ref{sec:resana} analyzes analytically the ionized multi-species coupling. We continue with  investigating the importance of this coupling in the solar atmosphere using  MHD models (Section~\ref{sec:revres}). Our research includes the following numerical models (Section~\ref{sec:res}): 1D tests to validate the implementation (Section~\ref{sec:1dnum}), and 1D numerical simulations  including momentum exchange to investigate the dissipation of high-frequency standing waves (Section~\ref{sec:1ddiss}), 1D numerical simulations of high-frequency Alfv\'en waves (Sections~\ref{sec:alf}). %and 2D reconnection (Section~\ref{sec:2drec}). 
This manuscript ends with a discussion and conclusions (Section~\ref{sec:con}). 

\section{Multi-fluid and Multi-species MHD Numerical Code}~\label{sec:methods}

We will dedicate a separate manuscript to a detailed description of the implementation and validation of the multi-fluid multi-species (MFMS) MHD equations used in \ebysus\ code. Therefore,  for the purpose of this article, we limit ourselves to a short description of the code and the governing equations. \ebysus\ has inherited the numerical methods used in the \bifrost\ code \citep{Gudiksen:2011qy}: Spatial derivatives and the interpolation of variables are done using sixth and fifth-order polynomials. As in \bifrost , the numerical scheme is defined on a staggered Cartesian mesh for which non-uniform grid spacing in one direction is allowed.  The equations are stepped forward in time using the modified explicit third-order predictor-corrector procedure \citep{hyman1979} allowing variations in time and operator splitting. Numerical noise is suppressed using high-order artificial diffusion. In contrast to the \bifrost\ code, which solves the radiative-MHD equations for a single fluid, the \ebysus\ code solves the MHD equations separately for each of the desired number of excited levels, ionization stages, and species as detailed below. In addition, \ebysus\  takes into account the electron momentum equation through the induction equation and the derivation of the electric field and, independently, the electron energy equation. 

For clarity and consistency we will use the same nomenclature as used by \citet{Ballester:2018fj} with minor adjustments. The ionization states are referred as $I$, \ie\ $I=0$ denotes neutrals and $\hat{I} = I \geq 1$ ions. The excited levels are marked with $E$ and the identity of the chemical species (or molecules) is indicated by $a$.
Consequently, each set of particles in a given micro-state will be described with $aIE$. 
For electrons the notation $aIE$ is reduced to just $e$. For simplicity, $\sum_\mathrm{a'}$ is the sum over all the species $a'$, $\sum_{I',a}$ is the sum over all ionization levels, including neutrals, for a given species $a$ and $\sum_{E',aI}$ is the sum over all the excited levels for a given ionized species $aI$. For clarity, we define $\sum_\mathrm{a'I'E'} = \sum_\mathrm{a'}\sum_{I',a'}\sum_{E',a'I'}$, and  $\sum_{I'E',a} = \sum_{I',a}\sum_{E',aI'}$.

\subsection{Continuity Equations}~\label{sec:cont}

The mass density for each type of species in a given micro-state is governed by the continuity equation in this generic form: 

\begin{eqnarray}
\frac{\partial \rho_\mathrm{aIE}}{\partial t} + \nabla \cdot \rho_\mathrm{aIE}\,  {\vec u_\mathrm{aIE}} = \sum_{I'E',a} \nonumber \\ m_\mathrm{aIE}(n_\mathrm{aI'E'}\Gamma^{ion}_\mathrm{aI'E'IE}-n_\mathrm{aIE}\Gamma^{rec}_\mathrm{aIEI'E'}) \label{eq:conti}  
\end{eqnarray} 

\noindent where $\rho_\mathrm{aIE}=m_\mathrm{aIE}\, n_\mathrm{aIE}$, ${\vec u_\mathrm{aIE}}$, $n_\mathrm{aIE}$ and $m_\mathrm{aIE}$ are the mass density,  velocity,  number density and  particle mass for a given micro-state. $\Gamma^{rec}_\mathrm{aIEI'E'}$, and $\Gamma^{ion}_\mathrm{aI'E'IE}$ are the transition rate coefficients between levels $I'E'$ and $IE$ due to recombination or de-excitation, and ionization or excitation, respectively. The ionization, recombination, excitation and de-excitation terms have been added and are solved by applying a Newton-Raphson implicit method at each grid cell individually. 

In this particular study, for simplicity, we neglect ionization and recombination. Our numerical experiments focus on the solar upper chromosphere and transition region, and the timescales modeled here are much shorter ($\leq 1$~s) than, for example, hydrogen ($\geq 10^2$~s) and helium ($10^3-10^5$~s) ionization and recombination timescales
\citep{Carlsson:1992kl,Carlsson:2002wl,Golding:2014fk}. Non-equilibrium ionization effects for many heavier atoms occur on timescales of about $10$-–$100$~s for typical density values for transition region and corona \citep{2010ApJ...718..583S}. A detailed description of the ionization-recombination implementation and results will be detailed in upcoming studies. 

Since electrons move so fast and their mass is negligible we ignore the continuity equation for electrons in the \ebysus\ code \ie\ we assume quasi-neutrality: $n_\mathrm{e}=\sum_\mathrm{aIE} n_\mathrm{aIE}\, Z_\mathrm{aI}$ where $Z_\mathrm{aI}$ is the ionized state. We denote the average velocity of the ionized species as ${\vec u}_c=\sum_\mathrm{a\hat{I}E}{\vec u}_\mathrm{aIE}$ and of the neutrals as  ${\vec u}_n=\sum_\mathrm{a0E}{\vec u}_\mathrm{a0E}$.  

\subsection{Momentum Equations}~\label{sec:mom}

The momentum equations depend on the ionization state under consideration. The momentum equations for each $aIE$ written in SI are included in the \ebysus\ code as follows:

\begin{eqnarray}
&& \frac{\partial (\rho_\mathrm{a\hat{I}E} {\vec u_\mathrm{a\hat{I}E}}) }{\partial t}+ \nabla \cdot (\rho_\mathrm{a\hat{I}E} {\vec u_\mathrm{a\hat{I}E}} {\vec u_\mathrm{a\hat{I}E}} -\hat{\tau}_\mathrm{a\hat{I}E})  =  \nonumber \\
&& - \nabla P_\mathrm{a\hat{I}E} + \rho_\mathrm{a\hat{I}E} {\vec g}+ n_\mathrm{a\hat{I}E} q_\mathrm{a\hat{I}} \left({\vec E} + {\vec u_\mathrm{a\hat{I}E}} \times {\vec B} \right) + \nonumber \\
&& \sum_{\hat{I'}E',a}(\Gamma_\mathrm{a\hat{I'}E'\hat{I}E}^{ion}m_\mathrm{a\hat{I}E}\vec{u}_\mathrm{a0E'}-\Gamma_\mathrm{a\hat{I}EI'E'}^{rec}m_\mathrm{a\hat{I}E}\vec{u}_\mathrm{a\hat{I}E}) \nonumber \\
&& +\sum_\mathrm{a'I'E'}\vec{R}_\mathrm{a\hat{I}E}^{a\hat{I}Ea'I'E'} \label{eq:spmoma1}\\
&& \frac{\partial (\rho_\mathrm{a0E} {\vec u}_\mathrm{a0E}) }{\partial t}+ \nabla \cdot (\rho {\vec u}_\mathrm{a0E} {\vec u}_\mathrm{a0E} -\hat{\tau}_\mathrm{a0E})  = \nonumber \\
&&  - \nabla P_\mathrm{a0E} + \rho_\mathrm{a0E} {\vec g} +\sum_\mathrm{a'I'E'}\vec{R}_\mathrm{a0E}^{a0Ea'I'E'}- \nonumber \\ 
&& \sum_{I'E',a}\left(\Gamma_\mathrm{aI'E'}^{ion}m_\mathrm{a0E}\vec{u}_\mathrm{a0E}+\Gamma_\mathrm{a0E}^{rec}m_\mathrm{aI'E'}\vec{u}_\mathrm{aI'E'}\right) 
\label{eq:spmomc1}
\end{eqnarray}

\noindent where $q_{\alpha}$, $P_{\alpha}$, and $\hat{\tau}_{\alpha}$ are the ion charge, gas pressure, and viscous tensor for a specific species (\ie\ $\alpha = aIE$).  ${\vec g}$, ${\vec E}$, and ${\vec B}$ are gravity acceleration, and electric and magnetic field, respectively. $\vec{R}_{\alpha}^{\alpha \beta}$ is the momentum exchange where $\alpha \neq \beta$ and both can be any $aIE$. We combined charge and momentum exchange, \ie\ we summed both cross sections \citep{Vranjes:2008uq}.
Due to the numerical stiffness, the momentum exchange terms are solved by operator splitting; implicitly using a Newton-Raphson method along with the ionization/recombination terms in the continuity equation discussed above. {\jms The momentum exchange can then be expressed as follows: 

\begin{eqnarray}
\vec{R}_{\alpha}^{\alpha \beta} = m_{\alpha}\, n_{\alpha}\, \nu_{\alpha \beta} ( \vec{u}_{\beta} - \vec{u}_{\alpha})  
\end{eqnarray}

\noindent where  $\nu_{\alpha \beta}$ is the collision frequency. Note that $\vec{R}_{\alpha}^{\alpha \beta} = - \vec{R}_{\beta}^{\beta \alpha } $. 

For neutral-ion and neutral-neutral, the collision frequencies ($\nu_{\alpha \beta}$) are typically expressed as follows:

\begin{eqnarray}
\nu_{\alpha \beta} = n_\beta \frac{m_\beta}{m_\alpha + m_\beta} C_{\alpha \beta}(T_{\alpha \beta}) 
\sqrt{\frac{8 K_B T_{\alpha \beta}}{\pi m_{\alpha \beta}}}
\end{eqnarray}

\noindent where the reduced temperature between the two fluids in K (see below) is $T_{\alpha \beta} = (m_\alpha T_\beta + m_\alpha T_\beta)/(m_\alpha + m_\beta)$ and $C_{\alpha \beta}(T_{\alpha \beta})$ is the cross section which is a function of temperature. Some relevant cross sections as a function of temperature are shown in \citep[][among others]{Krstic:1999rz, Glassgold:2005ty,Schultz:2008qv,Vranjes:2008uq,Vranjes:2013ve}.  We include both collisional momentum exchange and charge exchange. For those species where the cross section is not well-known we follow \citet{Vranjes:2008uq}, \ie\ the cross section between any other species with protons is chosen to be the value of the cross section for protons multiplied by $m_m/m_p$, where $m_m$ is the atomic mass of the considered species and $m_p$ is the proton mass.

For elastic ion-ion collisions, we consider Coulomb collisions following, for instance, \citet{Hansteen:1997ag}. The collision frequency is as follows:

\begin{eqnarray}
\nu_{a\hat{I}Ea'\hat{I'}E'} = 1.7 \frac{\ln \Lambda}{20} \frac{m_p}{m_{a\hat{I}E}} \left(\frac{m_{a\hat{I}Ea'\hat{I'}E'}}{m_p}\right)^{1/2} \nonumber \\
n_{a'\hat{I'}E'} T_{a\hat{I}Ea'\hat{I'}E'}^{-(3/2)} Z_{a'\hat{I'}}^2 Z_{a\hat{I}}^2
\end{eqnarray}

where $m_p$ is the proton mass, the collision frequency is measured in s$^{-1}$, and the ion number density in cm$^{-3}$, the reduced mass is   $m_{a\hat{I}Ea'\hat{I'}E'}=\frac{m_{a\hat{I}E}\,m_{a'\hat{I'}E'}}{m_{a\hat{I}E}+m_{a'\hat{I'}E'}}$, and the Coulomb logarithm is (all in SI units). 

\begin{eqnarray}
\ln \Lambda = 23.0+1.5\ln (T_e/10^6) - 0.5 \ln (n_e(cm^{-3})/10^6)
\end{eqnarray}

}

We ignore electron inertia and its time variation. We can then use the electron momentum equation to calculate the electric field (Section~\ref{sec:elect}). 

In addition to ionization and recombination, and to limit the scope of this study, to Isolate and highlight effects of velocity drift between ions we highly simplified our approach in this study  and we have also excluded the effects of gravity. 

\subsection{Energy Equations}~\label{sec:ener}

The internal energy equations for each $aIE$ implemented in the \ebysus\ code written in the SI conservative form are as follow:  

\begin{eqnarray}
&& \frac{\partial{e_\mathrm{a\hat{I}E}}}{\partial{t}} = -\nabla\cdot e_\mathrm{a\hat{I}E}\vec{u}_\mathrm{a\hat{I}E} -P_\mathrm{a\hat{I}E}\nabla\cdot\vec{u}_\mathrm{a\hat{I}E}  +\nonumber \\
&&  Q^{visc}_\mathrm{a\hat{I}E} + Q^{ir}_\mathrm{a\hat{I}E} + Q^{a\hat{I}Ee}_\mathrm{a\hat{I}E} + \sum_\mathrm{a'I'E'}^{a'I'E'\neq a\hat{I}E} Q_\mathrm{a\hat{I}E}^{a\hat{I}Ea'I'E'}+  \nonumber \\ 
&& q_\mathrm{a\hat{I}E} n_\mathrm{a\hat{I}E} {\vec u}_\mathrm{a\hat{I}E} \cdot \vec{E} \label{eq:enei5} \\ 
&& \frac{\partial{e_\mathrm{a0E}}}{\partial{t}} = -\nabla\cdot e_\mathrm{a0E}\vec{u}_\mathrm{a0E} -P_\mathrm{a0E}\nabla\cdot\vec{u}_\mathrm{a0E} + Q^{visc}_\mathrm{a0E} +\nonumber \\
&& Q^{ir}_\mathrm{a0E} + Q^{a0Ee}_\mathrm{a0E} + \sum_\mathrm{a'I'E'}^{a'I'E'\neq a0E} Q_\mathrm{a0E}^{a0Ea'I'E'} \label{eq:enen5} \\
&& \frac{\partial{e_{e}}}{\partial{t}} = -\nabla\cdot e_{e}\vec{u_{e}} -P_{e}\nabla\cdot\vec{u_{e}}  + Q^{visc}_e  +Q^{ir}_e + \nonumber \\
&& \sum_\mathrm{a'I'E'} Q_{e}^{ea'I'E'}+  q_{e} n_{e} {\vec u_{e}} \cdot \vec{E} + Q_\mathrm{spitz} \label{eq:enee5} 
\end{eqnarray}

\noindent where $e_\alpha$ is the internal energy for any $\alpha = [e,aIE]$. The last term in Equation~\ref{eq:enei5} and the second last term in Equation~\ref{eq:enee5} are the Joule heating. $Q^{visc}_{\alpha}$ is the viscous heating. $Q^{ir}_{\alpha}$ is the heating/cooling term due to the ionization, recombination excitation and/or de-excitation energy exchange.  $Q_{\alpha}^{\alpha\beta}$ is the heating of the fluid $\alpha$ due to collisions with fluid $\beta$ for any  $\alpha = [e,aIE]$ and $\beta = [e,a'I'E']$ {\jms as $Q_{\alpha}^{\alpha\beta} = \vec{R}^\alpha_{\alpha\beta} \cdot ( \vec{u}_\beta - \vec{u}_\alpha)$. $Q_{sptiz}$ is the heating term due to thermal condition along the magnetic field}. The latter is implemented using same scheme as in the \bifrost\ code. For simplicity and the purpose of the current paper, we do not take into account $Q_{spitz}$, $Q^{ir}_{\alpha}$, $Q_{e}^{ea'I'E'}$,  and $Q^{aIEe}_\mathrm{aIE}$. 

\subsection{Equation of State}\label{sec:eos}

We assume MFMS ideal gases: 

\begin{equation}
P_\mathrm{aIE}=n_\mathrm{aIE}k_B T_\mathrm{aIE} = (\gamma-1) e_\mathrm{aIE}
\end{equation}

\noindent \ie\ each fluid/species has its own temperature and no internal degrees of freedom. As mentioned above, for this study, we are not taking into account the energies of ionization, recombination, excitation and de-excitation. {\jms Similarly, we need an equation of state for electrons: 

\begin{equation}
P_e=n_e k_B T_e = (\gamma-1) e_e, 
\end{equation}

\noindent where we assume quasi-neutrality to compute $n_e$.}

\subsection{Electric Field and the Induction Equation}~\label{sec:elect}

The electric field can be defined in any reference system:

\begin{eqnarray}
\vec{E}  = \vec{E_{\alpha}} - \vec{u}_{\alpha}\times\vec{B} \label{eq:elecadv}
\end{eqnarray}

In \ebysus\ we use the laboratory frame of reference  -- which in the case of the Sun means a frame of reference relative to the local solar conditions -- and compute $\vec{E_{\alpha}}$ from the electron momentum equation ($\vec{E}_{e}$, i.e. electron frame of reference) which simplifies the implementation for the two-fluids, \ie\ ion and neutrals, or single fluid descriptions. Then $\vec{E}_{e}$ is: 

\begin{eqnarray}
\vec{E}_{e} = \frac{\nabla P_e}{n_\mathrm{e} q_\mathrm{e}} 
+\frac{\sum_\mathrm{aIE}\vec{R}_{e}^{eaIE}}{n_\mathrm{e} q_\mathrm{e}}  \label{eq:e_ohm} 
\end{eqnarray}

\noindent where all the inertia terms and the gravitational force acting on the electrons have been neglected. In addition, ionization/recombination is neglected, assuming the electron continuity equation is in the fast ionization/recombination limit. 

The evolution of the magnetic field is governed by Maxwell equations. In
\ebysus, we consider the induction equation:

\begin{eqnarray}
\frac{\partial{\vec{B}}}{\partial{t}} &=& - \nabla\times\vec{E}  \\ \nonumber
&=&\nabla\times \left(\vec{u_e} \times \vec{B} - \frac{\nabla P_e}{n_\mathrm{e} q_\mathrm{e}} - \frac{\sum_\mathrm{aIE}\vec{R}_{e}^{eaIE}}{n_\mathrm{e} q_\mathrm{e}}\right)
\end{eqnarray}

For simplicity and the purpose of this work, we do not take into account the electron pressure gradient term (the Biermann battery, the second term on the right hand side of the equation), nor the momentum exchange term between electrons and ion species (third term). Since we have several species, the electron velocity, thanks to the assumption of quasi-neutrality, reads as follows:  

\begin{eqnarray}
\vec{u}_{e} = \left(\sum_{\alpha} \frac{n_\alpha q_\alpha \vec{u}_{\alpha}}{n_\mathrm{e}\, q_\mathrm{e}}\right) - \frac{\vec{J}}{q_\mathrm{e}\, n_\mathrm{e}}, ~\label{eq:vele}
\end{eqnarray}

\noindent and $\vec{J} = (\nabla \times \vec{B})/\mu_0$.

\section{Analytical Analysis}~\label{sec:resana}

In this section, we describe analytically the coupling of ionized multi-species {\jms due to the electric field}. First, we briefly describe the coupling in the momentum equation (Section~\ref{sec:iondf}). Second, how this coupling adds stiffness to the numerical code (Section~\ref{sec:cfl}) and finally we present an analytical solution (Section~\ref{sec:1dana}) in order to test the code (Section~\ref{sec:1dnum}). 

\subsection{Coupling in the Momentum Equation for Ionized Species}~\label{sec:iondf}

The Lorentz force in a single fluid, or a two-fluids (ions and neutrals) description is ${\vec J} \times {\vec B}$. However, in full MFMS (Equation~\ref{eq:spmoma1}), the {\jms interaction of the Lorentz force with each ionized species} reads as follows:

\begin{eqnarray}
&& \frac{\partial \rho_\mathrm{a\hat{I}E} u_\mathrm{a\hat{I}E}}{\partial t} += - n_\mathrm{a\hat{I}E}\, q_\mathrm{a\hat{I}E}\nonumber \\ 
&& \left[\left(\sum_\mathrm{a'\hat{I'}E'} \frac{n_\mathrm{a'\hat{I'}E'} q_\mathrm{a'\hat{I'}E'} \vec{u}_\mathrm{a'\hat{I'}E'}}{n_\mathrm{e}\, q_\mathrm{e}}\right) - \right. \nonumber \\ 
&& \left.\frac{\vec{J}}{q_\mathrm{e}\, n_\mathrm{e}}-\vec{u}_\mathrm{a\hat{I}E} \right]\times \vec{B}  ~\label{eq:momvele}
\end{eqnarray}

\noindent \citep[\eg\ see][]{Cramer:2001hl}. 

{\jms Observe, if} $|\sum_\mathrm{a'\hat{I'}E'} n_\mathrm{a'\hat{I'}E'} q_\mathrm{a'\hat{I'}E'} \vec{u}_\mathrm{a'\hat{I'}E'} -n_\mathrm{e}\, q_\mathrm{e}\vec{u}_\mathrm{a\hat{I}E}| << |\vec{J}|$ for any ionized element, the expression above becomes $\vec{J}\times\vec{B}$, which brings us back to single fluid or two-fluids nomenclature. 

Note that the various momentum equations are coupled {\jms with the electric field as detailed with the Equations}~\ref{eq:momvele}. The evolution of the velocity of an ionized species depends on the velocity of the other species as long as $|\sum_\mathrm{a'\hat{I'}E'} n_\mathrm{a'\hat{I'}E'} q_\mathrm{a'\hat{I'}E'} \vec{u}_\mathrm{a'\hat{I'}E'} -n_\mathrm{e}\, q_\mathrm{e}\vec{u}_\mathrm{a\hat{I}E}|$ is large enough in relation to the other terms in the momentum equation. Collisions and ionization/recombination couple the momentum equations. We will focus on cases where  $|\sum_\mathrm{a'\hat{I'}E'} n_\mathrm{a'\hat{I'}E'} q_\mathrm{a'\hat{I'}E'} \vec{u}_\mathrm{a'\hat{I'}E'} -n_\mathrm{e}\, q_\mathrm{e}\vec{u}_\mathrm{a\hat{I}E}|$ indeed may becomes important. 

Let us define ion velocity drift {\em force} as follows: 

\begin{eqnarray}
&& \vec{F}^{IDrift}_{\mathrm{a\hat{I}E}}= - n_\mathrm{a\hat{I}E}\, q_\mathrm{a\hat{I}E}\nonumber \\ 
&& \left[\left(\sum_\mathrm{a'\hat{I'}E'} \frac{n_\mathrm{a'\hat{I'}E'} q_\mathrm{a'\hat{I'}E'} \vec{u}_\mathrm{a'\hat{I'}E'}}{n_\mathrm{e}\, q_\mathrm{e}}\right) - \vec{u}_\mathrm{a\hat{I}E} \right]\times \vec{B} \nonumber 
\\
&&= \frac{n_\mathrm{a\hat{I}E} q_\mathrm{a\hat{I}E}}{n_\mathrm{e}\, q_\mathrm{e}}\nonumber \\ 
&& \left[\sum_\mathrm{a'\hat{I'}E'\neq a\hat{I}E} n_\mathrm{a'\hat{I'}E'} q_\mathrm{a'\hat{I'}E'} (\vec{u}_\mathrm{a\hat{I}E} - \vec{u}_\mathrm{a'\hat{I'}E'})    \right]\times \vec{B}  ~\label{eq:ivdf2}  
\end{eqnarray}

\noindent \ie\ $\vec{J}\times \vec{B}$ has been subtracted. This term is not zero when two or more ionized species move at different velocities. Another interesting aspect is that this term does not appear in the induction equation since the electric field is a function of the bulk velocity of all ions (Equation~\ref{eq:vele}) and changes in the velocity due  to $F^{IDrift}$ will not impact the electric field. 

Because of the coupling due to the velocity drift force, ions can experience a synchronized  cyclotron {\jms type} motion. In other words, when several species are present {\jms with different velocities}, they will be coupled to the other ionized species and will experience {\jms (as shown in the next section) a rotational motion with a frequency that depends on} a combination of single-ion cyclotron-frequencies, the number density of ionized species and their ionization {\jms state \citep[as detailed below; similarly][derived this expression]{Martinez-Gomez:2016im}}. For two ionized species, this coupling leads to a \wc\ frequency: 

\begin{eqnarray} 
\hat{\Omega}_\mathrm{a\hat{I}Ea'\hat{I'}E'} = \frac{Z_\mathrm{a\hat{I}}  n_\mathrm{a\hat{I}E} \Omega_\mathrm{a'\hat{I'}E'} + Z_\mathrm{a'\hat{I'}}  n_\mathrm{a'\hat{I'}E'} \Omega_\mathrm{a\hat{I}E}}{n_{e}} ~\label{eq:wcyfr}
\end{eqnarray} 

\noindent where the cyclotron frequency for a specific species is $\Omega_\mathrm{a\hat{I}E} = \frac{q_\mathrm{a\hat{I}}  |B|}{m_\mathrm{a\hat{I}E}}$ (see Section~\ref{sec:1dana}). By increasing the number of ionized species, the number of superpositions of \wc\ frequencies increases. The \wc\ frequency becomes a combination of the cyclotron frequencies of the ionized species (Section \ref{sec:1dnum}).

\subsection{CFL}~\label{sec:cfl}

The ion velocity drift force adds a new Courant Friedrichs and Lewy (CFL) convergence condition \citep{Courant:1928uq} for MFMS MHD equations due to the \wc\ frequencies. Consequently, the length of a timestep cannot be greater than $1/\hat{\Omega}_\mathrm{a\hat{I}Ea'\hat{I}'E'}$, and this condition has to be fulfilled for each coupling between ionized species, considering the largest $\Omega_\mathrm{a\hat{I}Ea'\hat{I}'E'}$. 

\subsection{An Analytic Solution}~\label{sec:1dana}

In this section, following \citet{Cramer:2001hl}, we derive the analytic formulation of the \wc\ wave in simplified conditions without the effects of ionization and recombination, radiative losses, thermal conduction, the Hall term, Biermann battery, nor momentum exchange {\jms and gravity}. We use that solution to validate our numerical experiments.

In an atmosphere with no spatial variation in any of the thermal and density variables for any of the micro-states $aIE$, and given a constant magnetic field, \ie\ $\vec{J} = 0$, the multi-fluid momentum equations can be simplified to: 

\begin{eqnarray}
&& \frac{\partial \vec{u}_\mathrm{a\hat{I}E}}{\partial t} = \frac{ q_\mathrm{a\hat{I}E}}{m_\mathrm{a\hat{I}E} n_\mathrm{e}\, q_\mathrm{e}}\nonumber \\ 
&& \left[\sum_\mathrm{a'\hat{I'}E'\neq a\hat{I}E} n_\mathrm{a'\hat{I'}E'} q_\mathrm{a'\hat{I'}E'} (\vec{u}_\mathrm{a\hat{I}E}-\vec{u}_\mathrm{a'\hat{I'}E'})   \right]\times \vec{B}  ~\label{eq:ivdf2} 
\end{eqnarray}

The magnetic field does not change in time $\frac{\partial{\vec{B}}}{\partial{t}} = 0$ because we imposed $\vec{J} = 0$ and the bulk velocity and density are constant in space. This set of equations has as a solution a superposition of sinusoidals, as we show below. For simplicity, let us consider only two singly-ionized fluids, \eg\ $H^+$ and $He^+$. Since we allow only singly ionized species, \ie\ $q_\mathrm{e} = q_\mathrm{H^+} =q_\mathrm{He^+} $, and  $n_\mathrm{e} = n_\mathrm{H^+}+ n_\mathrm{He^+}$, then: 

\begin{eqnarray}
&& \frac{\partial\vec{u}_\mathrm{H^+}}{\partial t} =  \frac{q_\mathrm{e} n_\mathrm{He^+}}{m_\mathrm{H^+}}  \left[ \frac{\vec{u}_\mathrm{H^+} - \vec{u}_\mathrm{He^+}}{ n_\mathrm{H^+}+ n_\mathrm{He^+}}  \right]\times \vec{B}  ~\label{eq:sivdfH} \\
&& \frac{\partial \vec{u}_\mathrm{He^+}}{\partial t} =  \frac{q_\mathrm{e} n_\mathrm{H^+}}{m_\mathrm{He^+}}  \left[ \frac{ \vec{u}_\mathrm{He^+} - \vec{u}_\mathrm{H^+}}{n_\mathrm{H^+}+ n_\mathrm{He^+}} \right]\times \vec{B}  ~\label{eq:sivdfHe}
\end{eqnarray}

Note that the acceleration of a particular species depends on the particle mass of that species and on the number density of the other ionized species. The larger the number density from one species the larger the acceleration for the other species. These two equations can be combined in order to derive the evolution of the velocity drift ($\vec{u}_{D_\mathrm{H^+He^+}} = \vec{u}_\mathrm{H^+} - \vec{u}_\mathrm{He^+}$). 

\begin{eqnarray}
\frac{\partial\vec{u}_{D_\mathrm{H^+He^+}}}{\partial t} =  \frac{q_\mathrm{e}}{n_\mathrm{e}} \left[\frac{n_\mathrm{He^+}}{m_\mathrm{H^+}} \vec{u}_{D_\mathrm{H^+He^+}}  + \right. \\ \nonumber \left. \frac{n_\mathrm{H^+}}{m_\mathrm{He^+}} \vec{u}_{D_\mathrm{H^+He^+}} \right]\times \vec{B}  ~\label{eq:sivdfHe}
\end{eqnarray}

For a velocity drift perpendicular to the magnetic field (for instance, within the x-z plane {\jms and the magnetic field is aligned with the y-axis}) the equations can be reduced to: 

\begin{eqnarray}
&& \frac{\partial u_{Dx_\mathrm{H^+He^+}}}{\partial t} = - \frac{q_\mathrm{e}}{n_\mathrm{e}} \left[\frac{n_\mathrm{He^+}}{m_\mathrm{H^+}} + \frac{n_\mathrm{H^+}}{m_\mathrm{He^+}}  \right] u_{Dz_\mathrm{H^+He^+}} B_y  ~\label{eq:ivdfHx} \\
&& \frac{\partial u_{Dz_\mathrm{H^+He^+}}}{\partial t} =  \frac{q_\mathrm{e}}{n_\mathrm{e}} \left[\frac{n_\mathrm{He^+}}{m_\mathrm{H^+}}  + \frac{n_\mathrm{H^+}}{m_\mathrm{He^+}} \right] u_{Dx_\mathrm{H^+He^+}} B_y  ~\label{eq:ivdfHz}
\end{eqnarray}

One can decompose these equations into a wave-like form: 

\begin{eqnarray}
&& u_{Dx_\mathrm{H^+He^+}} = U^o_D \sin\left(\frac{q_\mathrm{e}}{n_\mathrm{e}} \left[\frac{n_\mathrm{He^+}}{m_\mathrm{H^+}} + \frac{n_\mathrm{H^+}}{m_\mathrm{He^+}} \right] B_y t \right) ~\label{eq:ivdfHx} \\
&& u_{Dz_\mathrm{H^+He^+}} = - U^o_D \cos\left(\frac{q_\mathrm{e}}{n_\mathrm{e}} \left[\frac{n_\mathrm{He^+}}{m_\mathrm{H^+}}  + \frac{n_\mathrm{H^+}}{m_\mathrm{He^+}} \right] B_y t \right) ~\label{eq:ivdfHz}
\end{eqnarray}

\noindent {\jms where $U_D^o$ is the absolute initial ion velocity drift}. This simple solution can be used to test the numerical implementation (Section~\ref{sec:1dnum}). The generic solution for many ionized species is a complex combination of velocity drifts and not Equation~\ref{eq:wcyfr}.
The number of superposition sinusoidal functions will increase with number of ionized levels and species and these frequencies depend on the atomic mass, magnetic field strength, charge and number density for each ionized micro-state and electron number density.  

\section{Importance of the Multi-species in the Solar Atmosphere}~\label{sec:revres}

In order to better understand how important multi-fluid and multi-species effects are in the solar atmosphere, we use in this section the time-dependent 2.5D radiative MHD numerical model by \citet{Martinez-Sykora:2019hhegol} calculated with the \bifrost\ code. This model solves the hydrogen and helium rate equations including non-equilibrium ionization \citep{Leenaarts+etal2007,Golding:2016wq} and ion-neutral interaction effects introduced through Generalized Ohm's Law \citep{Martinez-Sykora:2012uq,Martinez-Sykora:2017gol,Nobrega-Siverio:2020arXiv200411927N}. The numerical domain spans from the upper layers of the convection zone, through the photosphere, chromosphere and transition region, to the corona. The advantage of using this model is that it aims to replicate more realistic conditions than we can for now with the \ebysus\ code, thus allowing insight into typical parameters of the solar atmosphere that can be used for \ebysus\ studies.
Figure~\ref{fig:2dtgrb} shows temperature, density and magnetic field strength maps, revealing low lying transition loops, magneto-acoustic shocks as well as two types of spicules, and including a million degrees corona. The magnetic field configuration is designed to mimic a plage region \citep[for further details, see][]{Martinez-Sykora:2019hhegol}. 

\begin{figure}[tbh]
	\begin{center}
		\includegraphics[width=0.99\hsize]{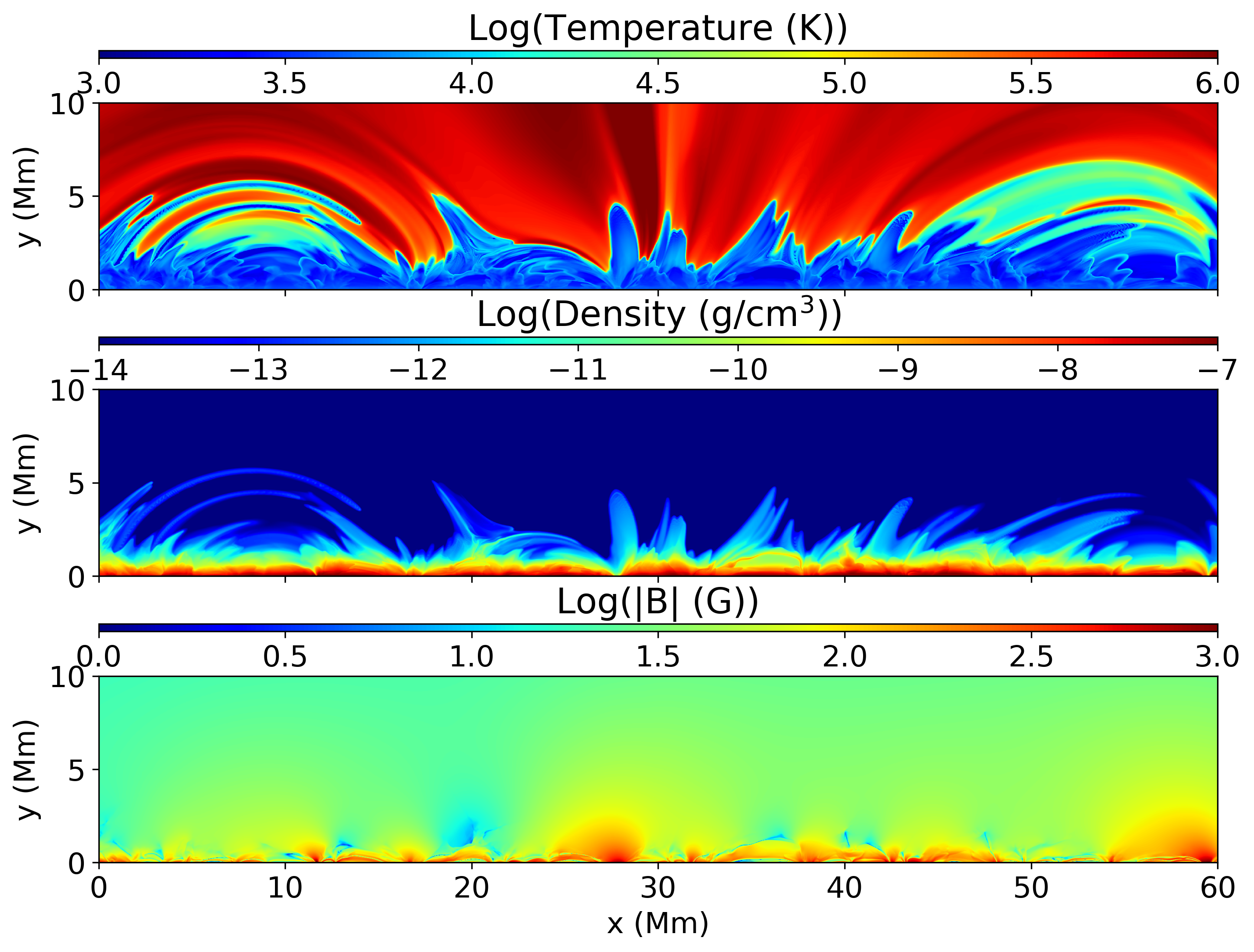}
		\caption{We use \citet{Martinez-Sykora:2019hhegol}'s 2.5D radiative MHD simulation to investigate the importance of the ion cyclotron resonance described in the previous section. From top to bottom, temperature, density and absolute magnetic field strength maps are shown for the selected instant in the simulated timeseries. \label{fig:2dtgrb}}
	\end{center}
\end{figure}

\subsection{From a Collisional to a Collisionless Multi-species Solar Atmosphere}~\label{sec:atm}

The solar atmosphere has a highly complex chemical composition \cite[\eg][]{Asplund:2009uq}. As long as the collision frequencies between species which constitute solar plasma are slow compared with other relevant physical processes, multi-species interaction effects must be taken into account. 

\begin{figure}[tbh]
	\begin{center}
		\includegraphics[width=0.99\hsize]{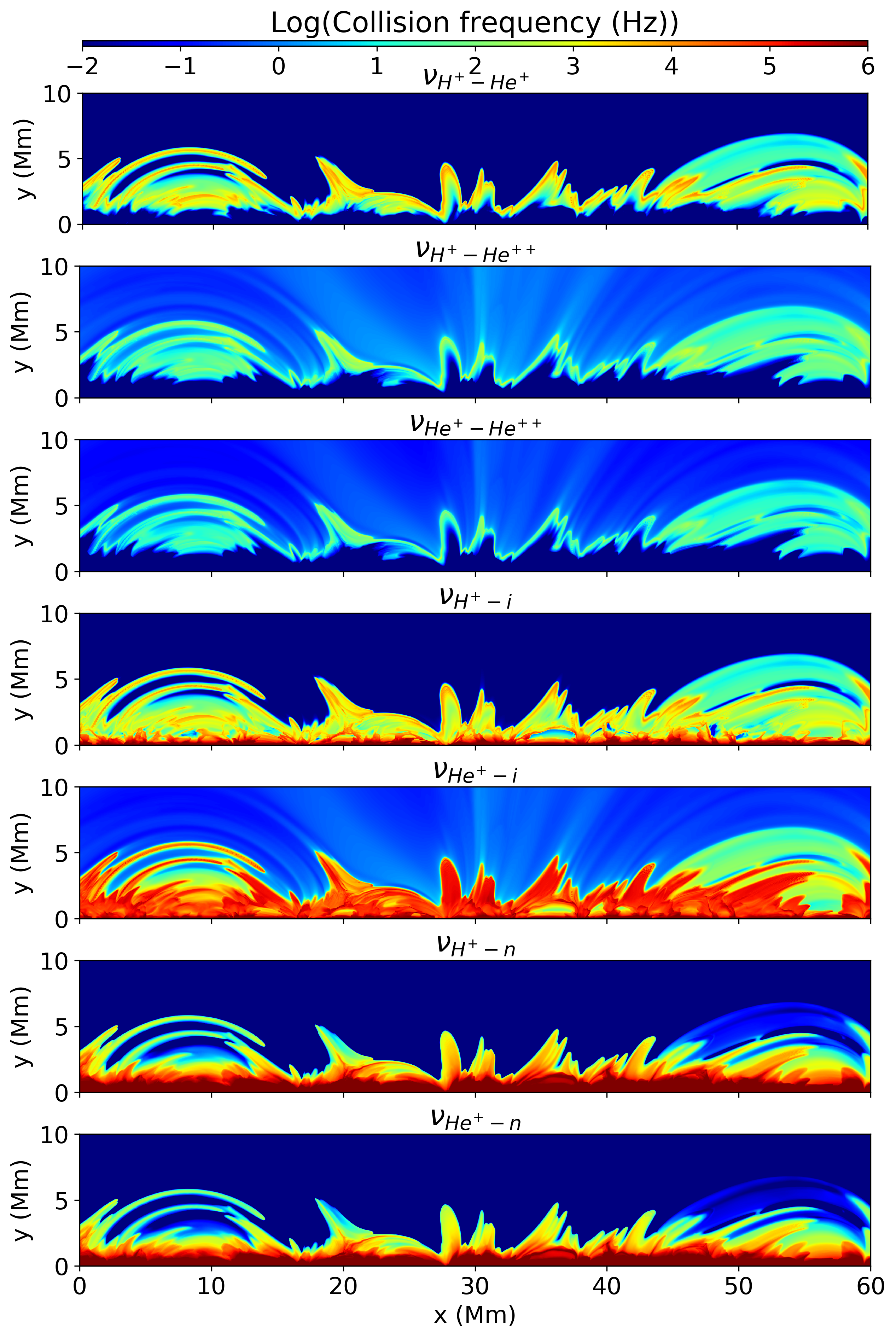}
		\caption{The collision frequency between different ionized species is small in the upper chromosphere, transition region and corona.  \hp-\hep, \hp-\hepp, \hep-\hepp, \hp-i, \hep-i, \hp-n, and \hep-n collision frequencies are shown from top to bottom. Ions (i) and neutrals (n) include the sixteen top most abundant species.}\label{fig:2dcol} 
	\end{center}
\end{figure}

We have estimated various collision frequencies from \citet{Martinez-Sykora:2019hhegol}'s time-dependent 2.5D radiative MHD simulation. \hp-\hep, \hp-\hepp, \hep-\hepp, \hp-i, \hep-i, \hp-n, and \hep-n collision frequencies are shown in Figure~\ref{fig:2dcol} from top to bottom respectively. 
The estimated collision frequencies between \hp\ or \hep\ with ions in the upper chromosphere and transition region ranges from a few 1~Hz to $10^3$~Hz. For the processes of interest in this paper, we will show that thermo-dynamic timescales can be shorter and the multi-species interaction effects can become relevant in upper chromosphere, TR and lower corona. 

\subsection{Weighted Averaged Ion-cyclotron Frequencies and Collisions: Upper Chromosphere and TR}~\label{sec:waf_coll_atm}

\begin{figure*}[tbh]
	\begin{center}
	    \includegraphics[width=0.99\hsize]{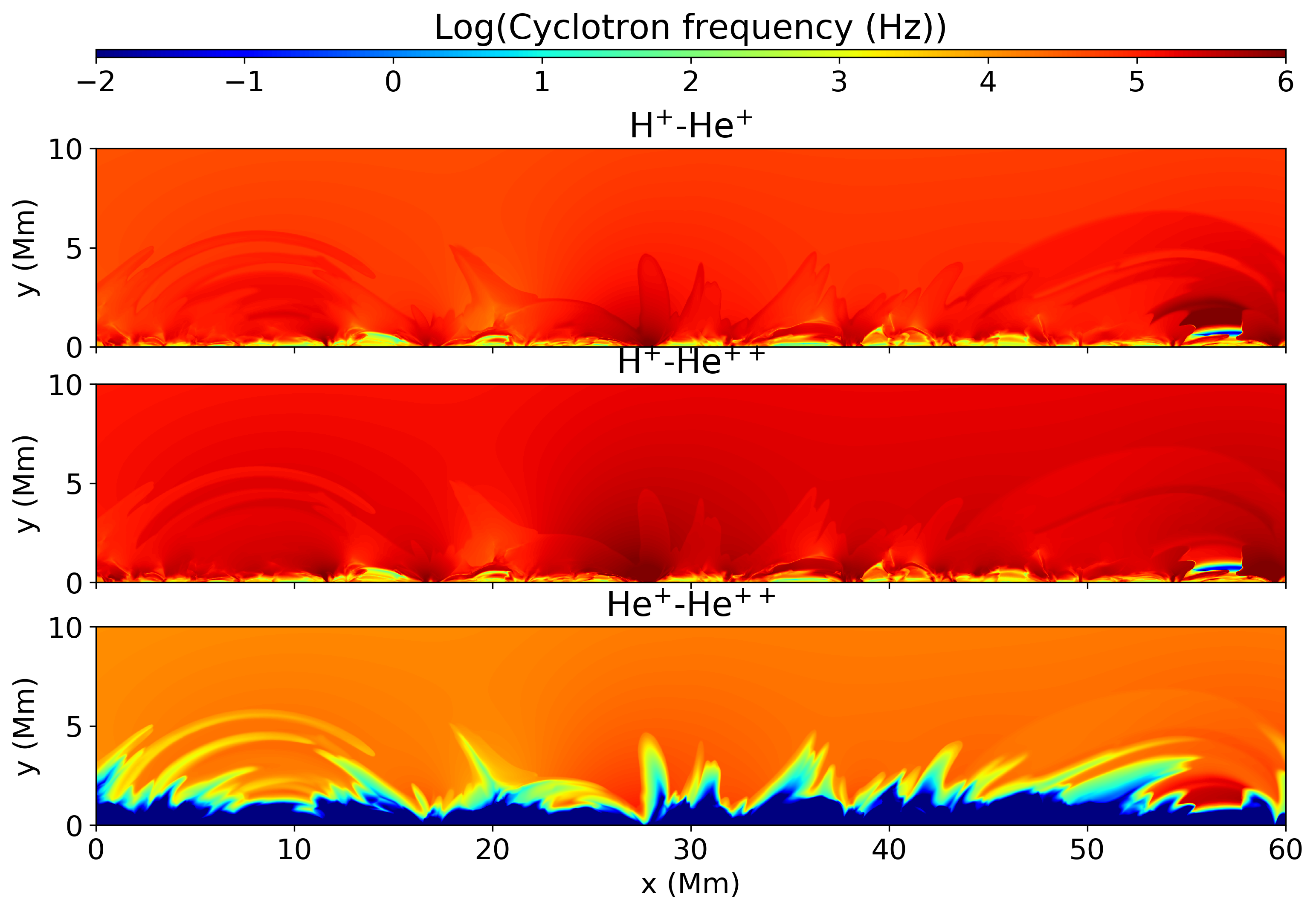}
		\caption{The \wc\ frequencies frequency is high in the upper chromosphere and corona. The \wc\ frequency between ionized hydrogen and singly ionized helium level,  ionized hydrogen and double ionized helium level, and singly ionized helium level and double ionized helium level are shown from top to bottom. \label{fig:2dcyc}}
	\end{center}
\end{figure*}

It is thus of great interest to compare the collisions frequencies between different ionized species and the \wc\ frequencies.
Using the \citet{Martinez-Sykora:2019hhegol} 2.5D radiative MHD simulation, we also calculated the \wc\ frequencies between \hp-\hep, \hp-\hepp, and \hep-\hepp\ using expression~\ref{eq:wcyfr}  (from top to bottom in Figure~\ref{fig:2dcyc}). For \hp-\hep\ and \hp-\hepp, these \wc\ frequencies are high, ranging from $10^2$ to $10^6$ Hz. Within the chromosphere, they become smaller (ranging from $10$ to $10^4$~Hz) in cold regions (yellow areas in top two panels near z=0.5~Mm). The lowest \wc\ frequencies are for \hep-\hepp\ (from $1$ to $10^4$ Hz). 

The velocity drift between the different ionized species, or the \wc\ wave amplitudes, can be reduced by collisions between species. Taking into account the analysis from the previous section, the collision frequency between different ionized species is very small in the upper chromosphere, transition region and corona. Typical collisional chromospheric timescales are of the order of a fraction of second (which frequency is $\simeq 1$ Hz). Figure~\ref{fig:2dcycrat} compares collision frequencies with \wc\ frequencies. In the lower chromosphere and below, the \wc\ frequency is much lower than the collision frequencies. Consequently, collisions will not allow the production of ion-cyclotron waves. In the corona, collisions are much less frequent and thus cannot significantly alter these waves. From the mid-chromosphere to the {\jms transition region} \wc\ waves can become important for interactions between \hp-\hep\ and \hp-\hepp, and for \hep-\hepp\ in the transition region. {\jms These regions of interest includes spicules, low laying loops and dynamic fibrils.} In these regions where the collisions are comparable to the \wc\ frequencies, these waves can be generated, propagate over a certain distance, and be dissipated through collisions. In Section~\ref{sec:alf} we show that Alfv\'en waves can generate ion-velocity drifts and this means that collisions can provide an efficient mechanism for dissipation of high-frequency Alfv\'en waves \citep{Martinez-Gomez:2016im}.

\begin{figure}[tbh]
	\begin{center}
		\includegraphics[width=0.99\hsize]{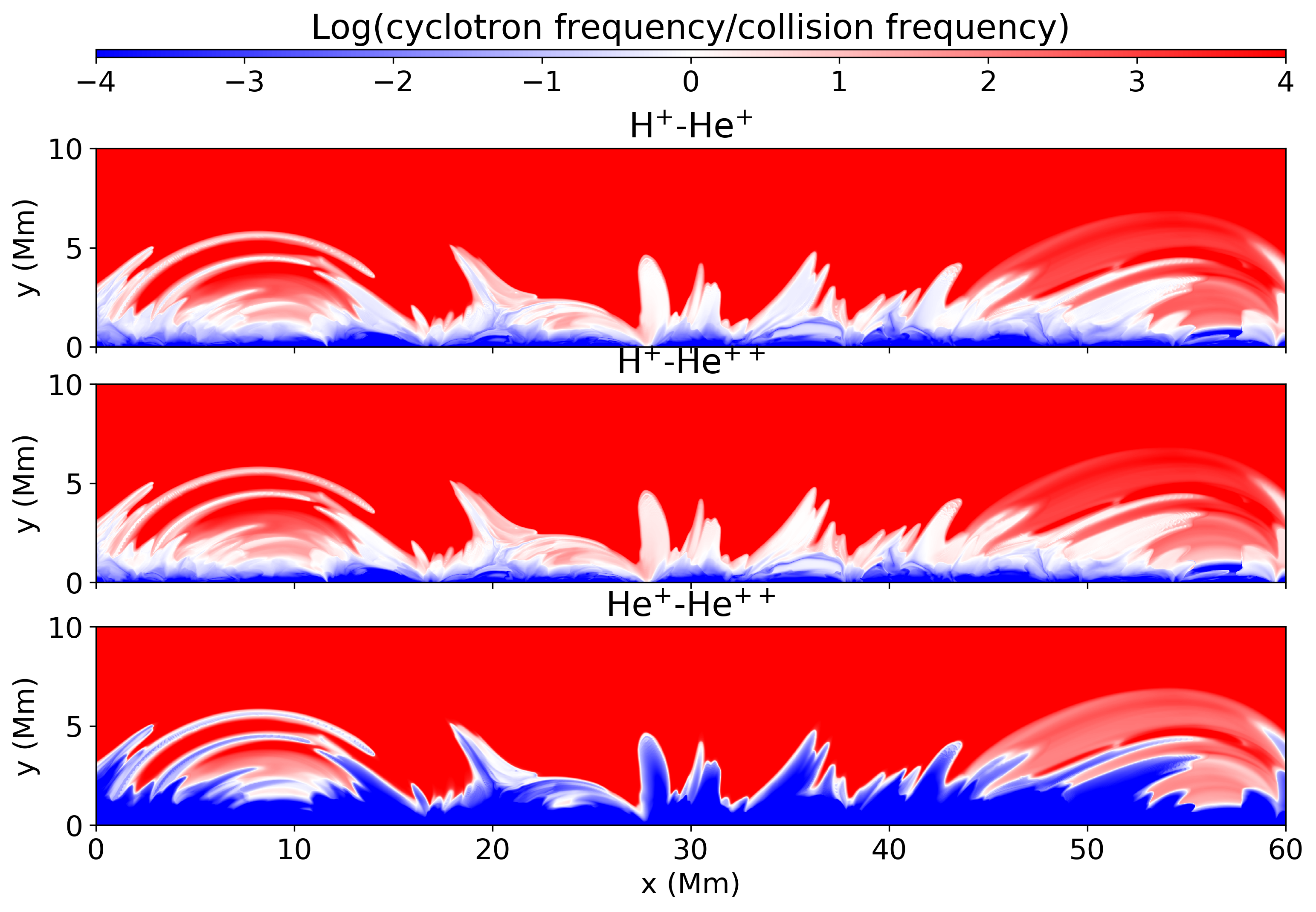}
		\caption{The \wc\ waves will facilitate the dissipation of Alfv\'enic waves in the chromosphere and TR. Ratio between these cyclotron frequencies with the collision frequencies between ionized hydrogen and first ionized helium level, ionized hydrogen and second ionized helium level, and first ionized helium level and second ionized helium level are shown from top to bottom.\label{fig:2dcycrat}}
	\end{center}
\end{figure}

In the single fluid description, even when including ion-neutral interaction effects (\eg\ ambipolar or Pedersen diffusion), or for a two-fluid description such as ions and neutrals or ions and electrons, we impose the same velocity for all species, and all ionized species ($a\hat{I}E$) move at the same speed. The same is true for all neutral species ($a0E$). Consequently, these simplified descriptions implicitly assume that 1) forces acting on each $a\hat{I}E$ and on each $a0E$ should be the same 2) the collisional time-scales must be small enough 3) and drivers must act on time-scales larger than the \wc\ frequencies, \ie\ velocity drifts between species must be negligible compared to bulk ionized plasma velocity. 
However, as we show in the above, in the solar atmosphere these conditions likely do not always apply. For example, the forces acting on each $aIE$ are not always the same, since each $aIE$ may have a different mass and ionization state. In addition, as we have shown in this section, the collision frequency between different ionized species is very small in the upper chromosphere, transition region and corona, and therefore the velocity drift between ionized species may not be damped by collisions. 

\section{Numerical Experiments}~\label{sec:res}

Of all the various physical processes that \ebysus\ is capable of handling, we here only use a subset. The main subject of the current work is to study and isolate the electric force ($F^{IDrift}$) due to ion drift velocities, which is the main subject of this work. As mentioned above, neither gravity, ionization and recombination, the Hall term, the Biermann battery, radiative losses, nor thermal conduction are taken into account here. 

First, we analyze three sets of 1D numerical simulations (Section~\ref{sec:1dnum}). The first set of 1D numerical tests is used to compare the codes' implementation with the analytic solution described in Section~\ref{sec:1dana}. The second set of 1D numerical models focuses on the possible role of the ion-velocity drift in the upper chromospheric thermodynamics (Section~\ref{sec:1ddiss}). The last set of 1D numerical models is focused on \wc\ waves driven by Alfv\'en waves (Section~\ref{sec:alf}). 
%We complete this section with a collisionless MFMS 2D reconnection test (Section~\ref{sec:2drec}).  
Table~\ref{tab:1ds} summarizes these experiments with the list of species included in each of the experiments along with relevant initial and/or boundary conditions.

\begin{table*}
	\centering
	\caption{List of  numerical simulations. From left to right: list the name, type of species and number of ionized levels and other properties of the simulated domain.}\label{tab:1ds}	
	\begin{tabular}{|l|l|l|}
		\hline
		Name & $aI$ & Other properties \\ \hline  \hline
		1D2S2LNI & H:2 lvl, He:2 lvl & Collisionless and initial ion velocity drift \\ \hline
		1D3S2LNI & H:2 lvl, He:2 lvl, C:2 lvl &  Collisionless and initial ion velocity drift \\ \hline
		1D2S2LCI & H:2 lvl, He:2 lvl & Collisional and initial ion velocity drift \\ \hline
		1D2S2LB0 & H:2 lvl, He:2 lvl & Collisionless \& boundary driven ($0.1$~Hz) \\ \hline
		1D2S2LB1 & H:2 lvl, He:2 lvl & Collisionless \& boundary driven ($10$~Hz, $\Delta B_x=0.25$~G) \\ \hline
		1D2S2LB2 & H:2 lvl, He:2 lvl & Collisionless \& boundary driven ($20$~Hz, $\Delta B_x=0.25$~G)\\ \hline
		1D2S2LB3 & H:2 lvl, He:2 lvl & Collisionless \& boundary driven ($10^{2}$~Hz, $\Delta B_x=0.25$~G)\\ \hline
		1D2S2LB4 & H:2 lvl, He:2 lvl & Collisionless \& boundary driven ($200$~Hz, $\Delta B_x=0.25$~G)\\ \hline
		1D2S2LB5 & H:2 lvl, He:2 lvl & Collisionless \& boundary driven ($10^{3}$~Hz, $\Delta B_x=0.25$~G)\\ \hline
    	1D2S2LB6 & H:2 lvl, He:2 lvl & Collisionless \& boundary driven ($10^{5}$~Hz, $\Delta B_x=0.25$~G)\\ \hline
		1D2S2LB7 & H:2 lvl, He:2 lvl & Collisionless \& boundary driven ($10^{3}$~Hz, $\Delta B_x=0.025$~G)\\ \hline
		1D2S2LB8 & H:2 lvl, He:2 lvl & Collisionless \& boundary driven ($10^{3}$~Hz, $\Delta B_x=2.5$~G)\\ \hline
		%2D1S0L & H:1 lvl (ions) & Single fluid 2D reconnection simulation\\ \hline
		%2D2S2LNR & H:2 lvl, He:2 lvl & Collisionless \& 2D reconnection simulation\\ \hline
	\end{tabular}
\end{table*}

\subsection{1D Collisionless Numerical Experiments}~\label{sec:1dnum}

The 1.5D collisionless numerical experiments of \wc\ standing waves shown here validates the numerical implementation used in \ebysus\ by comparison with the analytical solutions described in Section~\ref{sec:1dana}. 

The numerical domain in all 1.5D numerical experiments shown here are set up along the $z$-axis and cover the range $z=[0,1]$~Mm with an uniform grid of 200 points. Boundaries are periodic. The constant magnetic field is perpendicular to the simulated numerical domain, and is oriented along the $y$-axis. We impose an initial velocity along the $z$-axis, perpendicular to the magnetic field. Note that the ion-cyclotron frequency is independent of the internal energy and the thermodynamic properties of neutral species (see Equation~\ref{eq:wcyfr}).

\begin{sidewaystable*}[ht]
    \centering
    \vspace{-8cm}
	\begin{tabular}{|l|l|l|l|l|l|l|l|l|l|l|l|l|}
		\hline
		ID&$\rho_\mathrm{H^0}$&$n_\mathrm{H^0}$&$\rho_\mathrm{H^+}$&$n_\mathrm{H^+}$&$\rho_\mathrm{He^0}$&$n_\mathrm{He^0}$&$\rho_\mathrm{He^+}$&$n_\mathrm{He^+}$&$\rho_\mathrm{C^0}$&$n_\mathrm{C^0}$&$\rho_\mathrm{C^+}$&$n_\mathrm{H^0}$ \\ \hline \hline
		1D2S2LNI&$1.4\,10^{-10}$& $8.5\, 10^{13}$& $5.8\, 10^{-10}$ & $3.5\,10^{14}$ & $1.4\, 10^{-10}$ & $2.2\,10^{13}$ & $1.4\, 10^{-10}$ & $2.2\,10^{13}$ & N/A & N/A & N/A & N/A \\ \hline
		1D3S2LNI&$3\,10^{-9}$&$1.8\, 10^{15}$&$4\,10^{-9}$&$2.4\,10^{15}$&$1.4\,10^{-9}$ &$2.1\,10^{14}$&$1.4\,10^{-9}$&$2.1\,10^{14}$&$6.6\,10^{-11}$&$3.3\,10^{12}$& $2.6\,10^{-9}$ & $1.3\,10^{14}$\\ \hline
		1D2S2LCI&$4\,10^{-17}$ &$2.7\,10^{7}$&$7.3\,10^{-14}$&$4.4\,10^{10}$& $6.5\,10^{-15}$&$0.8\,10^8$&$1.8\,10^{-14}$&$2.8\,10^{9}$& N/A & N/A & N/A & N/A\\ \hline
		1D2S2LB0&$2\,10^{-19}$ &$1.3\,10^{5}$&$7.2\,10^{-13}$&$4.3\,10^{11}$& $2.9\,10^{-15}$&$4.3\,10^8$&$2.9\,10^{-13}$&$4.3\,10^{10}$& N/A & N/A & N/A & N/A\\ \hline
		1D2S2LB1&$2\,10^{-19}$ &$1.3\,10^{5}$&$7.2\,10^{-13}$&$4.3\,10^{11}$& $2.9\,10^{-15}$&$4.3\,10^8$&$2.9\,10^{-13}$&$4.3\,10^{10}$& N/A & N/A & N/A & N/A\\ \hline
		1D2S2LB2&$2\,10^{-19}$ &$1.3\,10^{5}$&$7.2\,10^{-13}$&$4.3\,10^{11}$& $2.9\,10^{-15}$&$4.3\,10^8$&$2.9\,10^{-13}$&$4.3\,10^{10}$& N/A & N/A & N/A & N/A\\ \hline
		1D2S2LB3&$2\,10^{-19}$ &$1.3\,10^{5}$&$7.2\,10^{-13}$&$4.3\,10^{11}$& $2.9\,10^{-15}$&$4.3\,10^8$&$2.9\,10^{-13}$&$4.3\,10^{10}$& N/A & N/A & N/A & N/A\\ \hline
		1D2S2LB4&$2\,10^{-19}$ &$1.3\,10^{5}$&$7.2\,10^{-13}$&$4.3\,10^{11}$& $2.9\,10^{-15}$&$4.3\,10^8$&$2.9\,10^{-13}$&$4.3\,10^{10}$& N/A & N/A & N/A & N/A\\ \hline
		1D2S2LB5&$2\,10^{-19}$ &$1.3\,10^{5}$&$7.2\,10^{-13}$&$4.3\,10^{11}$& $2.9\,10^{-15}$&$4.3\,10^8$&$2.9\,10^{-13}$&$4.3\,10^{10}$& N/A & N/A & N/A & N/A\\ \hline
		1D2S2LB6&$2\,10^{-19}$ &$1.3\,10^{5}$&$7.2\,10^{-13}$&$4.3\,10^{11}$& $2.9\,10^{-15}$&$4.3\,10^8$&$2.9\,10^{-13}$&$4.3\,10^{10}$& N/A & N/A & N/A & N/A\\ \hline
		1D2S2LB7&$2\,10^{-19}$ &$1.3\,10^{5}$&$7.2\,10^{-13}$&$4.3\,10^{11}$& $2.9\,10^{-15}$&$4.3\,10^8$&$2.9\,10^{-13}$&$4.3\,10^{10}$& N/A & N/A & N/A & N/A\\ \hline
		1D2S2LB8&$2\,10^{-19}$ &$1.3\,10^{5}$&$7.2\,10^{-13}$&$4.3\,10^{11}$& $2.9\,10^{-15}$&$4.3\,10^8$&$2.9\,10^{-13}$&$4.3\,10^{10}$& N/A & N/A & N/A & N/A\\ \hline
	%\end{tabular}	
	%\begin{tabular}{|l|l|l|l|l|l|l|l|l|}
		\hline
		ID&$\vec{u_{\mathrm{H^+}}}$&$\vec{u_{\mathrm{He^+}}}$&$\vec{u_\mathrm{C^+}}$ & $T$ & $\vec{B}$ & $v_a$ & dz & nz &&&&  \\ \hline  \hline
	    1D2S2LNI&$(-3.6,0,34)$ &$(14,0,-200)$ & N/A & $10^4$ &$(0,1,0)$ & $0.02$& $5$ & 200 &&&&\\ \hline
	    1D3S2LNI&$(-2.9,0,4.5)$ &$(5.7,0,-20)$&$1.4,-15$&$10^4$&$(0,1,0)$&$0.08$& $5$ & 200 &&&&\\ \hline
		1D2S2LCI&$(-4.3,0,0.04)$ &$(-9.3,0,2.2)$ & N/A &  $4\, 10^4$ &	$(0,10,0)$&$90$& $100$ & 200 &&&&\\ \hline
		1D2S2LB0&(0,0,0) & (0,0,0) & N/A &  $4\, 10^4$ & $(0,0,10)$ &$28$& $10^{-2}$ & 2000 &&&&\\ \hline
		1D2S2LB1&(0,0,0) & (0,0,0) & N/A &  $4\, 10^4$ & $(0,0,10)$ &$28$&  $1.2\,10^{-2}$ & 2000 &&&&\\ \hline
		1D2S2LB2&(0,0,0) & (0,0,0) & N/A &  $4\, 10^4$ & $(0,0,10)$ &$28$& $6\,10^{-3}$ & 2000 &&&&\\ \hline
		1D2S2LB3&(0,0,0) & (0,0,0) & N/A &  $4\, 10^4$ & $(0,0,10)$ &$28$& $2\,10^{-3}$ & 2000 &&&& \\ \hline
		1D2S2LB4&(0,0,0) & (0,0,0) & N/A &  $4\, 10^4$ & $(0,0,10)$ &$28$& $10^{-3}$ & 2000 &&&&\\ \hline
		1D2S2LB5&(0,0,0) & (0,0,0) & N/A &  $4\, 10^4$ & $(0,0,10)$ &$28$& $2\,10^{-4}$ & 2000 &&&&\\ \hline
		1D2S2LB6&(0,0,0) & (0,0,0) & N/A &  $4\, 10^4$ & $(0,0,10)$ &$28$& $3\,10^{-6}$ & 2000  &&&&\\ \hline
		1D2S2LB7&(0,0,0) & (0,0,0) & N/A &  $4\, 10^4$ & $(0,0,10)$ &$28$& $2\,10^{-4}$ & 2000 &&&&\\ \hline
		1D2S2LB8&(0,0,0) & (0,0,0) & N/A &  $4\, 10^4$ & $(0,0,10)$ &$28$& $2\,10^{-4}$ & 2000 &&&&\\ \hline
	\end{tabular}
\caption{\label{tab:1dt} Initial thermodynamic properties of the 1D numerical simulations. Left to right: mass density(g~cm$^{-3}$), number density (cm$^{-3}$), initial velocity (km~s$^{-1}$) for \hp, \hep, and \cp , temperature (K), magnetic field (G), Alfv\'en speed (km~s$^{-1}$), grid spacing (km) and number of grid points. All experiments have velocities 0~km~s$^{-1}$ for neutrals.}	
\end{sidewaystable*}

Figure~\ref{fig:1dmulti} shows the various velocity components for each ionized species as a function of time for experiment 1D2S2LNI. This experiment includes two species, \ie\ hydrogen and helium, with two levels each: neutrals and single ionized ions (Table~\ref{tab:1ds}). The magneto-thermal properties are constant with a magnetic field strength {\jms of $B_y=1$~G}. Note that these preliminary models are meant to validate the implementation and we choose low magnetic field strengths in order to avoid prohibitively small timesteps. Other initial thermodynamic properties of the simulations are listed in Table~\ref{tab:1dt}. Figure~\ref{fig:1dmulti} shows that the cyclotron motion is coupled between ionized hydrogen and ionized helium due to the electric field (Equation~\ref{eq:momvele}). The analytic solution (Equation~\ref{eq:wcyfr}) for 1D2S2LNI provides a frequency of $62$~Hz, for all ionized species and velocity components in agreement with the numerical experiments. The different components ($x$ and $z$) of the velocities of ionized species are 90 degrees out of phase, i.e., they rotate. In addition, the velocities of the two ionized species are in anti-phase as expected from Equations~\ref{eq:sivdfH}-\ref{eq:ivdfHz}. The velocity amplitude is larger for He$^{+}$ because, as mentioned above, the acceleration depends on its own particle mass and the number of particles of the other species. Since He$^{+}$ weighs more than H$^{+}$, this  reduces the accelerations in He$^{+}$, but since $n_{H^+}$ is much larger than $n_{He^+}$, the acceleration is larger in He$^+$ than in H$^+$.

\begin{figure}[tbh]
	\begin{center}
		\includegraphics[width=0.95\hsize]{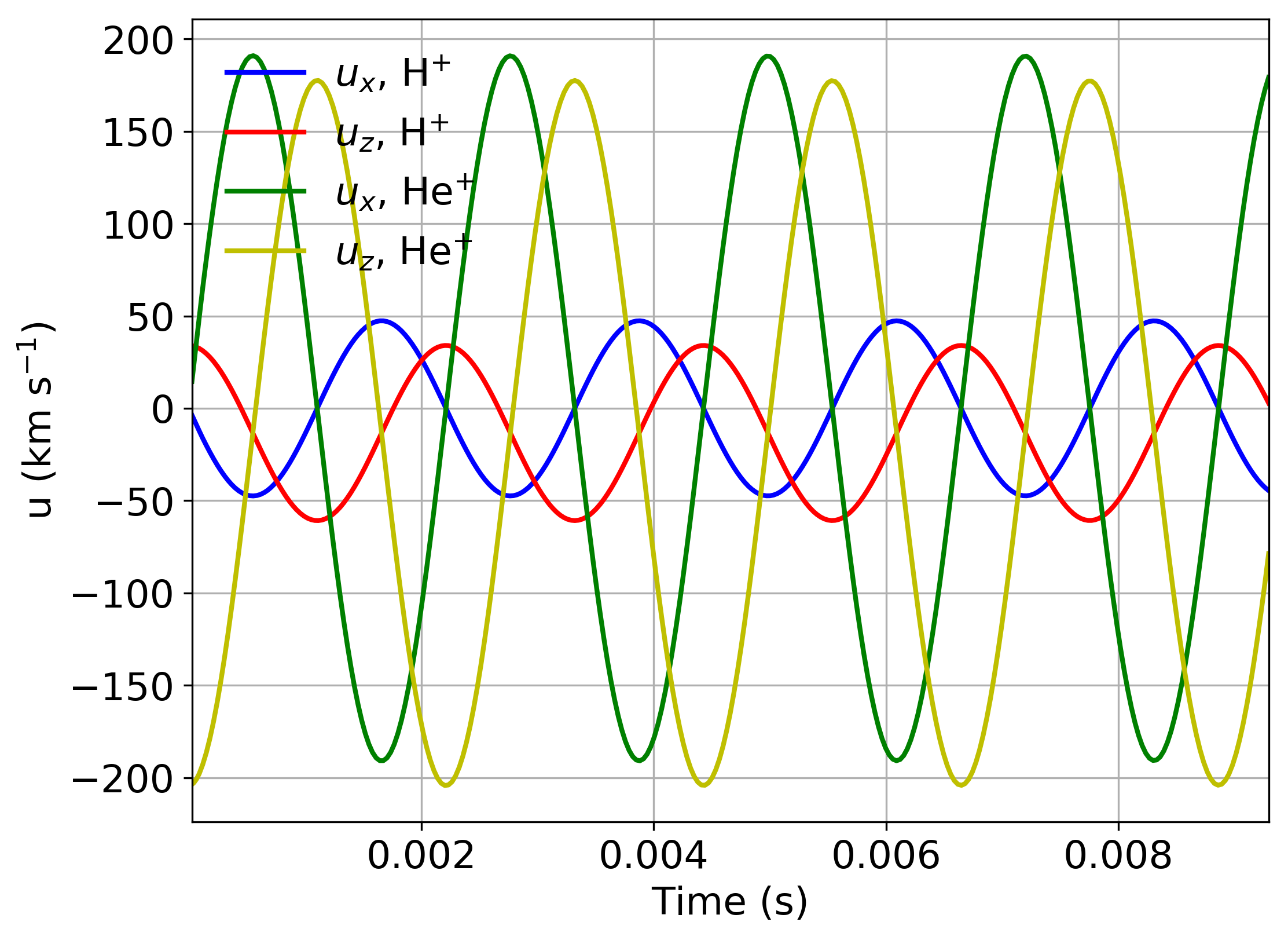}
		\caption{Note that the various components of the velocity for any ionized species and levels have the same frequency. Component $x$ and $z$ of the velocity as a function of time for the various ionized species (see labels) are shown for simulation 1D2S2LNI. \label{fig:1dmulti}}
	\end{center}
\end{figure}

As mentioned above, adding new ionized species will add wave superposition with different frequencies and amplitudes. Note that Equation~\ref{eq:wcyfr} is only valid for two ionized species. Figure~\ref{fig:1dmulti3} shows the various velocity components for the ionized species of simulation 1D3S2LNI which includes three species (hydrogen, helium and carbon) with two levels each (Table~\ref{tab:1ds}).  Table 2 provides densities for each species and levels. H and He follow photospheric abundances. In order to have enough ionized C, we increased the density and, therefore, the abundance with respect to the photospheric abundance in order to illustrate complex coupling since the various \wc\ depend on the density number of each ionized species (\eg\ Equations~\ref{eq:ivdfHx}-\ref{eq:ivdfHz}). Otherwise, since the amplitude on these multi-ion frequencies dependent on the density, C contribution will be negligible. The \wc\ frequency is not as trivial to derive as for two ionized species (Section~\ref{sec:1dana}) and becomes a combination of the cyclotron frequency of the various ionized species. With three different ionized species one can discern two different \wc\ frequencies in Figure~\ref{fig:1dmulti3}. In short, number densities, atomic mass, field strength and charge play a role in the amplitudes and frequencies as described above and following Equations~\ref{eq:sivdfH}-\ref{eq:ivdfHz} \citep[see also][]{Martinez-Gomez:2016im}.

\begin{figure}[tbh]
	\begin{center}
		\includegraphics[width=0.95\hsize]{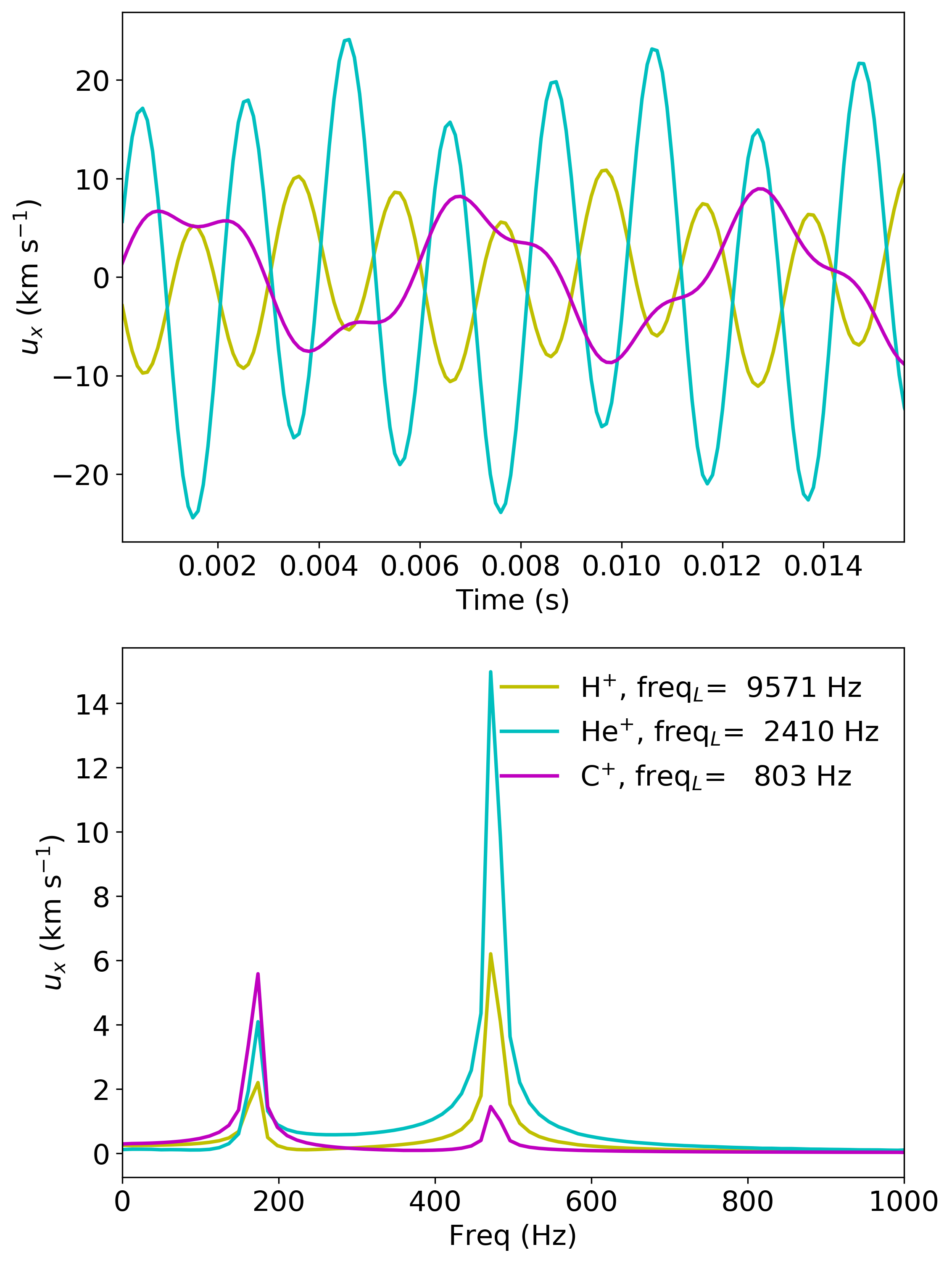}
		\caption{Component $x$ of the velocity as a function of time (top) and frequency (bottom) for the various species, hydrogen, helium, and carbon, for their ionized level (see labels) for simulation 1D3S2LNI. {\jms For comparison, we have added the relevant Lamor frequencies to the labels.} \label{fig:1dmulti3}}
	\end{center}
\end{figure}

\subsection{Kinetic Energy Dissipation}~\label{sec:1ddiss}

Momentum exchange between species could dissipate any \wc\ waves that may occur in the upper chromosphere, TR and lower corona  (Figure~\ref{fig:2dcyc}). Let us investigate if such dissipation can heat the solar atmosphere. Experiment 1D2S2LCI has typical upper-chromosphere/transition region densities and energies (Table~\ref{tab:1dt}). The simulated atmosphere is uniform and includes hydrogen and helium with two levels each (neutrals and first ionization level, Table~\ref{tab:1ds}). This numerical experiment includes momentum exchange. The top and middle panels of Figure~\ref{fig:1ddamping} show the $x$ and $z$ components of the velocity, respectively, and the bottom panel the temperature for each level of each species. The amplitude of the \wc\ wave is damped {\jms by a factor $e$ due to momentum exchange in roughly $3\,10^{-4}$~s}. Note that neutrals also show temperature variations in time due to the momentum transfer from the ions. This dissipation heats the corresponding level of each species following: 

\begin{eqnarray}
Q^{\alpha \beta}_{\alpha} = \vec{R^{\alpha \beta}_\alpha} \cdot (\vec{u_\alpha}-\vec{u_\beta})
\end{eqnarray}

Since \hep\ initially has the largest amplitude and the lowest solar number densities, it reaches the highest temperatures. Due to collisions, the plasma is not only heated but the temperatures of all species converge  following: 

\begin{eqnarray}
Q^{\alpha \beta}_{\alpha} = 3 \frac{m_{\alpha \beta}}{m_\alpha} n_\alpha \nu_{\alpha \beta} k_B (T_\beta-T_\alpha)
\end{eqnarray}

So, in this case the helium is heated first through dissipation of the drift velocities, and then helium heats the other species and fluids. The time needed for the different species of the plasma to reach similar temperatures is larger than the time needed to damp the \wc\ wave and this depends on the heating as well the density. {\jms The coupling between the different fluids is different. For instance, neutral hydrogen is more coupled to the ions than neutral helium is due to their different cross-sections as well as the densities of the different fluids (Figure~\ref{fig:2dcol})}.  For the lower densities of the upper transition region or corona, the damping times-scales increase by an order of magnitude or more.

\begin{figure}[tbh]
	\begin{center}
		\includegraphics[width=0.92\hsize]{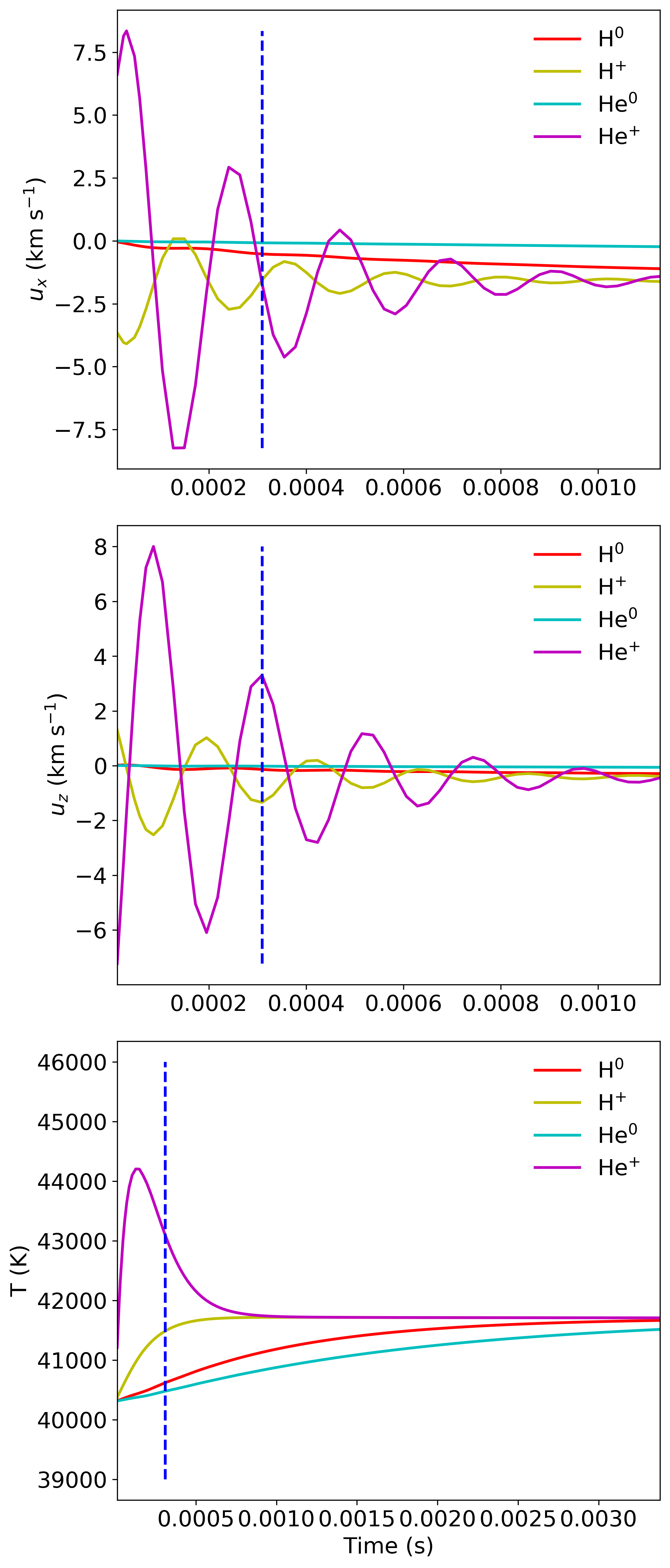}
		\caption{The \wc\ wave amplitude is damped due to momentum exchange. Component $x$ (top) and $z$ (middle) of the velocity and temperature (bottom panel) as a function of time for the various energy levels for each species are shown for simulation 1D2S2LCI. Note that the time range of the velocity plots is a subset of the time range shown in the bottom panel  (as indicated by the blue line between the bottom and panels and with dashed blue line in the bottom panel). \label{fig:1ddamping}}
	\end{center}
\end{figure}

Consequently, if these \wc\ waves (or velocity drift between ionized species) can be generated, this may be an effective dissipation and heating mechanism for the plasma in the upper chromosphere and TR and lower corona. Therefore, it is of great interest to investigate which mechanisms could lead to the formation of such waves and how much energy is contained in such high-frequency modes. Current MHD models do not include the physics of these drift velocities, neither their generation or their dissipation. 

\subsection{Alfv\'en Waves}~\label{sec:alf}  

The equations and analysis in detailed in Section~\ref{sec:1dana} suggest that magnetic tension, such as the restoring force in Alfv\'en waves, may lead to ion velocity drifts.  \citet{Martinez-Gomez:2016im} analytical analysis from multi-fluid and multi-species studies show that in order to generate waves at the \wc\ frequencies, the Alfv\'en frequency  must be the same or greater than those frequencies. We performed seven 1D simulations driving  Alfv\'en waves of differing frequencies at the bottom boundary (Table~\ref{tab:1ds}). For the 1D2S2LB0, 1D2S2LB1, 1D2S2LB2, 1D2S2LB3, 1D2S2LB4, 1D2S2LB5, and 1D2S2LB6 simulations Alfv\'en waves are driven at frequencies of $0.1$ $10$, $20$, $10^{2}$, $200$, $10^{3}$ and $10^{5}$~Hz, respectively, which except for simulation 1D2S2LB6 are all smaller than the \wc\ frequency ($1.5\, 10^4$~Hz). In all these experiments, in order to drive the Alfv\'en wave, we impose a background magnetic field with an initial $B_z=10$~G. .  In order to drive the Alfven wave, we impose a temporally varying component with an  amplitude of  $B_x$ of 0.25~G, \ie\ perpendicular to the 1.5D numerical domain. This produced a velocity amplitude of the ions of 800~m~s$^{-1}$ (see the second row of Figure~\ref{fig:1dwave}) {\jms propagating along the $z$-axis.  These waves are transverse, but since we are in a 1.5D geometry, they behave as Alfv\'en waves.} We introduced a small amplitude in $B_x$ to mitigate numerical constraints on the time-step (but see below). For 1D2S2LB7, 1D2S2LB8 simulations Alfv\'en waves are driven at frequencies of $10^{3}$~Hz and we impose an amplitude of $B_x$ of 0.025~G and 2.5~G respectively, producing a velocity amplitude of the ions of 80~m~s$^{-1}$ and 8~km~s$^{-1}$, respectively.

\begin{figure*}[tbh]
	\begin{center}
		\includegraphics[width=0.99\hsize]{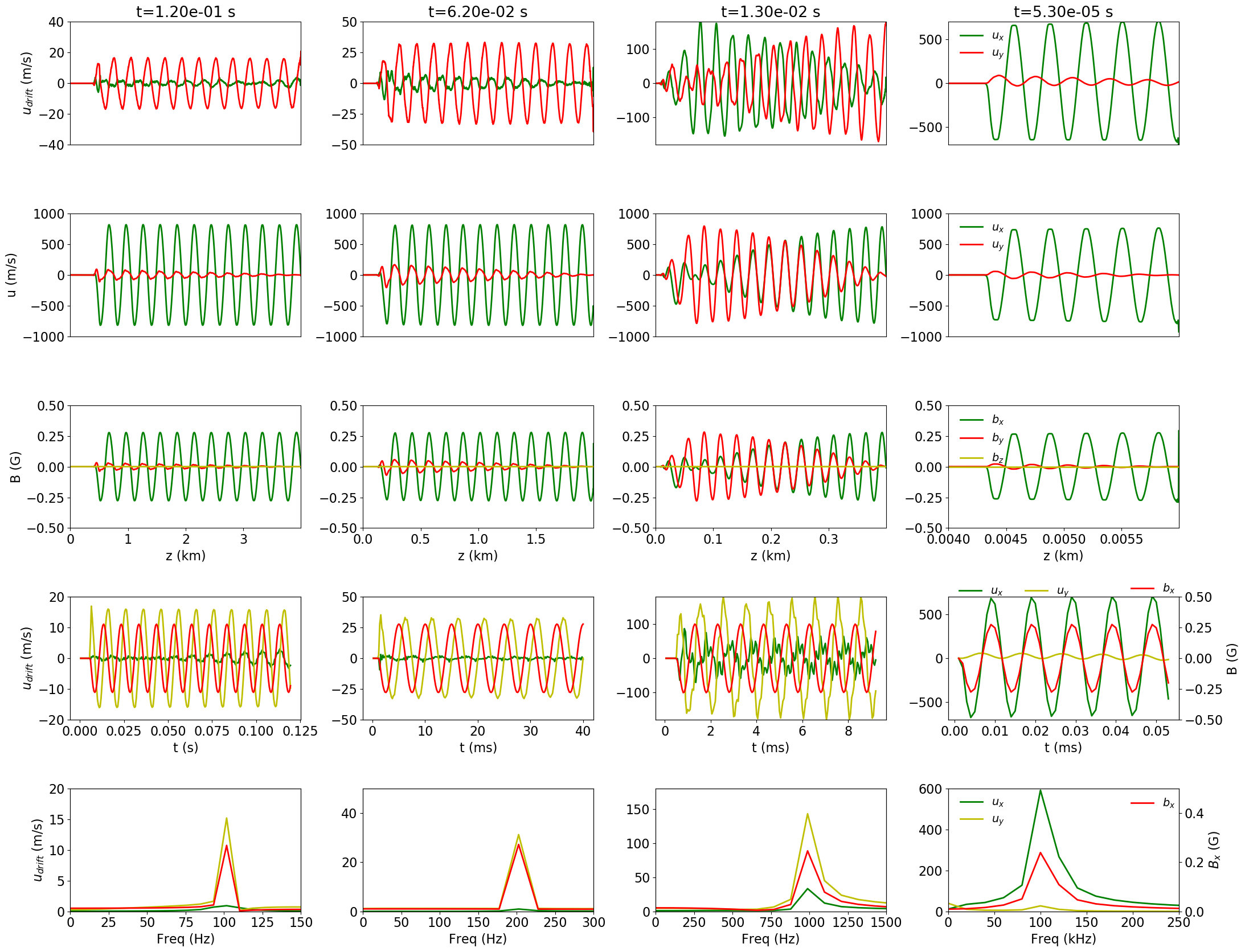}
		\caption{A velocity drift wave is driven by high-frequency Alfv\'en waves. The velocity drift between the two ionized species as a function of space are shown in the top row, from left to right, for the 1D2S2LB3, 1D2S2LB4, 1D2S2LB5, and 1D2S2LB6 simulations. The total velocity for all ions is shown in the second row. The magnetic field is shown in the third row. We subtracted the initial (10~G) component z of the magnetic field. Fourth and last rows show the velocity ($x$=green and $y$=yellow) and $x$ component of the magnetic field (red) as a function of time and frequency, respectively.  \label{fig:1dwave}}
	\end{center}
\end{figure*}

Figure~\ref{fig:1dwave} shows the velocity drift between the different ionized species (first, {\jms fourth} and last rows) and the magnetic field (second to last rows) for 1D2S2LB3, 1D2S2LB4, 1D2S2LB5, and 1D2S2LB6 simulations (left to right columns, respectively). Even for  driven Alfv\'en wave frequencies smaller than the \wc\ frequency (D2S2LB3, 1D2S2LB4, and 1D2S2LB5) a  velocity drift between the two ionized fluids appears in the component perpendicular to the background magnetic field and to the imposed Alfv\'en wave plane, \ie\ $y$. The velocity drift results from the fact that the inertia of and the forces acting on various ionized species are different.

The velocity drift frequencies match the imposed Alfv\'en wave frequency in all experiments. Therefore, the frequency of the velocity drift does not depend on \wc\ waves as was the case in the previous experiments in previous sections where we imposed an initial velocity drift.

The absolute velocity drift increases with the driven Alfv\'en wave frequency as long as these frequencies are lower than \wc\ frequencies. In addition, the velocity field changes components (\ie\ from $u_y$ to $u_x$) when the driven Alfv\'en wave frequency is greater than \wc\ frequency. In other words, for frequencies greater than \wc\ frequencies (1D2S2LB6), time is too short to produce the coupling between the ionized species and the current (and $\vec{J}\times \vec{B}$ in the momentum equation) dominates. Therefore, the velocity drift component corresponds to the  $\vec{J}\times \vec{B}$ component. The Lorentz force leads to a velocity drift due to the different inertia, number of ionized particles and charge between the different ionized species (see Equation~\ref{eq:momvele}). 

As the waves propagate in space, the components of velocity and magnetic field disturbance are transposed to the other component perpendicular to the direction of propagation. This other orthogonal component ($y$ for magnetic field, $x$ for velocity) increases with space in an oscillatory fashion. The length scale (along the propagation direction, $z$) over which this transposition from $x$ to $y$ (or $y$ to $x$) occurs depends on the driven Alfv\'en frequency: it decreases for increasing frequency. In other words, the higher the frequency, the shorter the distance over which the transfer from x to y occurs.

Figure~\ref{fig:1dwavefrevsamb} shows the linear dependence of the component $y$ of the velocity drift with the driver frequency. Simulations 1D2S2LB5, 1D2S2LB7 and 1D2S2LB8 show an increase/decrease of the ion velocity drift with the same factor as the increase/decrease of the driving amplitude. Therefore, for a driven frequency of $10^3$~Hz, $\Delta B_x=2.5$~G, and $B_z=10$~G the velocity drift is $1$~km~s$^{-1}$ and the amplitude of the ion velocity wave is 8~km~s$^{-1}$. The velocity drift appears even for frequencies lower than the \wc frequency.  We also run experiments where we maintained the $|B|/|\Delta B_x|$ ratio, but reduce/increase by an equal amount both $|B|$ and $|\Delta B_x|$. In these cases, for the same imposed Alfv\'en frequency, the amplitude of the ion velocity drift is the same. In other words, for the same $|B|/|\Delta B_x|$ ratio, the ion velocity drift is equal, be it in a sunspot or in internetwork.

The velocity drift increases linearly with the imposed Alfv\'en frequency.  Observational evidence for Alfv\'en waves in the solar chromosphere suggests that there is significant power at periods of 3 to 5 min. For such low frequencies, the wave energy in the velocity drift is likely negligible. However, some observations have found evidence for higher frequency waves with periods as short as $45$~s ($0.15$~Hz) in spicules \citep[\eg][]{Okamoto:2011kx}. These observations are limited by the cadence of the observations, so that even higher frequencies cannot be excluded by observations. In fact, transition region spectral lines often show significant broadening beyond the thermal width of order 20 km~s$^{-1}$ in exposure times as short as $4$~s \citep{De-Pontieu:2015dz}. If this non-thermal broadening were to be caused by waves, wave frequencies could be significantly higher than $1$~Hz.
For the modest Alfv\'en amplitudes studied here ($\Delta B_x=2.5$~G) in a $10$~G field, this would lead to velocity drifts of order $2$~m~s$^{-1}$, \ie\ quite low. However, these waves may continuously drive the ion velocity drift while they travel through the solar atmosphere and therefore will be damped as well. 

{\jms A zero-order calculation of the dissipated magnetic energy could be done as follows. Assuming these waves are dissipated while traveling from the mid-chromosphere to the transition region at the top of a spicule, \ie\ a distance of roughly $4$~Mm, (Figure~\ref{fig:2dcycrat}) and considering an Alfv\'en speed of 400~km~s~$^{-1}$ \citep{DePontieu:2017net}, it takes them roughly 10~s to propagate along the spicule. Alfv\'en waves such as the ones mentioned above, i.e., 10~G guided field, 2.5~G Alfv\'en amplitudes (equivalent to $\sim 10$~km~s$^{-1}$ wave velocity amplitudes) and a frequency of 1~Hz, will continuously drive a velocity drift of $\sim 2$~m~s$^{-1}$. Assuming the collisions damp this velocity drift in $\sim0.01$~s (top panel of Figure~\ref{fig:2dcol}), the total Alfv\'en amplitude damped through the chromosphere is 2~km~s$^{-1}$. For a spicule with $\sim 10^{-12}$~g~cm$^{-3}$ mass density (middle panel of Figure~\ref{fig:2dtgrb}), $\sim300$~km wide, and for a length of 4~Mm, the energy dissipated in the spicule is $\sim3\,10^{18}$~erg which is $\sim3\,10^{17}$~erg~s$^{-1}$, \ie\ an energy flux of  $\sim3\,10^{6}$~erg~s$^{-1}$~cm$^{-2}$ \citep[compare with the energy flux needed to maintain the corona][]{De-Pontieu:2007bd}. However, this is a back of envelope calculation, and the next step remains to} investigate it further in models that contain a stratified atmospheres, \ie\ more representative of what is expected to be the situation in the solar atmosphere, and in models where the parameter range for magnetic field strengths and driving wave amplitudes are extended. In addition, momentum exchange between species should also be included. 

\begin{figure}[tbh]
	\begin{center}
		\includegraphics[width=0.99\hsize]{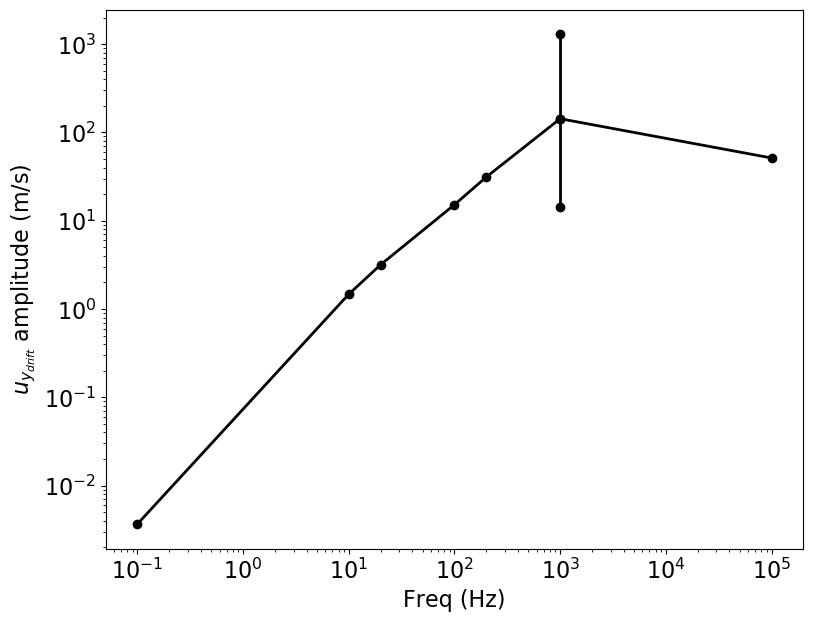}
		\caption{The velocity drift wave amplitude has a linear dependence with the frequency of the driven Alfv\'en waves for frequencies lower than the \wc . The top and bottom dots at $10^3$~Hz correspond to experiments 1D2S2LB7 and 1D2S2LB8, respectively. \label{fig:1dwavefrevsamb}}
	\end{center}
\end{figure}

\section{Conclusions and Discussion}~\label{sec:con}

In a weakly collisional environment, ionized species are coupled through the electric field. So, cyclotron motions occur at \wc\ frequencies, which depend on the weighted mass and charge of the ionized species. In addition, the Lorentz force and inertia can act differently on different species. We have developed a new numerical MHD code, called \ebysus,  that allows us to treat species and their ionized and excited levels separately. This code inherited many features from the \bifrost\ code and shares several numerical characteristics of that code. \ebysus\ allows us to run experiments of interest for this study on multi-ionized species interactions in the solar atmosphere.

In this investigation, we first describe  the theory behind multi-ion interactions. This has an analytical solution which is useful for testing the numerical code \citep[\eg][]{Cramer:2001hl}. An imposed ion-velocity drift can drive \wc\ waves. The more species the more superposed \wc\ frequencies. These frequencies depend on the number density, cyclotron frequency for each individual ionized species, and particle mass. Then, from so-called realistic radiative MHD models \citep{Martinez-Sykora:2019hhegol}, we estimated the \wc\ frequencies and collision frequencies between different species in the solar atmosphere. We find that these effects may play a role in the upper-chromosphere, TR, and corona.  Finally, we calculated highly idealized MFMS numerical experiments in order to understand the coupling between the ionized species through the electric field and its possible role in the solar atmosphere. 

The velocity drift between ionized species naturally occurs via Alfv\'en waves. %and especially in reconnection processes. 
In principle, as long as there is a significant magnetic tension{\jms, e.g. reconnection}, species will experience different forces due to the differences in charge and inertia. {\jms Preliminary results of magnetic reconnection reveal the presence of ion drift at different stages of the reconnection.}

As long as Alfv\'en waves frequencies are smaller than the \wc\ frequencies, the ion-velocity drift increases with increasing Alfv\'en frequencies. In addition, the ion velocity drift increases with wave amplitude and background magnetic field. Further, the ion velocity drift in invariant for a fix $|B|/|\Delta B_x|$, \ie\ in those cases the ion velocity drift is the same in a sunspot or in internetwork. The ion velocity drift might be minor, but there are, at least, three main points to consider in order to estimate if this process can dissipate the wave energy into thermal energy. First, the amplitude of the waves is not well known, with current observations suggesting amplitudes of, on average, about $20$~km~s$^{-1}$. Secondly, the frequency of the observed waves is poorly constrained, with some observations finding evidence for waves of 0.15 Hz on spicules \citep{Okamoto:2011kx}, and measurements of non-thermal broadening compatible (in principle) with frequencies higher than $1.6$~Hz \citep{De-Pontieu:2015dz}. Perhaps most importantly, these ion velocity drifts might be continuously generated as long as the Alfv\'en wave is propagating while at the same time these velocity drifts are being damped by collisions.  Therefore, in order to estimate the dissipation of  Alfv\'en waves in the solar atmosphere it is crucial to extend this research by studying models of a stratified atmosphere that includes collisions and a more realistic range of values for various other parameters. {\jms High-frequency waves could be driven by magnetic reconnection in the chromosphere \citep{Lazarian:1999ApJ...517..700L}. Similarly, the formation of type II spicules is also associated with high-frequency waves \citep{Okamoto:2011kx,Martinez-Sykora:2017sci}.}

%The presence of many ionized species in a collisionless reconnection model is shown to lead to fractionation and strong velocity drifts in different regions within the reconnection site. New waves, perturbations, and instabilities appear with MFMS effects. The chosen parameters for this simulation are typical for UV-Bursts, but collisions may also play a big role in these events and have been ignored here. For lower densities, i.e., in transition region reconnection events, collisions may become less important, so that the importance of MFMS effects may increase. A natural next step is to include collisions and reproduce other reconnection scenarios that could occur higher in the solar atmosphere.

The ion-velocity drift between ionized species may play a role from the mid-chromosphere to the corona. In addition, these drifts can be dissipated in the chromosphere and transition region and damp high-frequency Alfv\'en waves  and thus contribute in heating spicules and dynamic fibrils \citep{Martinez-Sykora:2019hhegol}.  \citet{chintzoglou2020iris} suggested that the simulated spicules may miss extra heating mechanisms. The proposed heating mechanism described here can be also important in the lower corona and it may provide a way to dissipate Alfvenic waves generated by spicules and/or reconnection events in the corona, and heat the associated loops \citep{Martinez-Sykora:2017rb,De-Pontieu:2017pcd,Henriques:2016ApJ...820..124H}. 

The Alfv\'en driven models %and the reconnection simulation 
have ion-drifts that can lead to a chemical fractionation. The scope of this study is limited and we have not analyzed this possibility any further. Alfv\'en driven models have (not shown here) ion velocity drift in the direction of the propagating wave (z-axis). One of the future goals of \ebysus\ is to investigate the First Ionization Potential (FIP) effect and chemical fractionation in the solar atmosphere. Our preliminary results indicate that this code is suited for this research.

This is a very first work on \wc\  and ion-drift velocities using highly simplified MFMS numerical simulations. The natural next step is to add complexity to the experiments to address the role of this effect and quantify how important it can be in the solar atmosphere. 

\section{Acknowledgments}

We gratefully acknowledge support by NASA grants, NNX17AD33G, 80NSSC18K1285, 80NSSC20K1272 and NNG09FA40C (IRIS), NSF grant AST1714955. The simulations have been run on clusters from the Notur project, and the Pleiades cluster through the computing project s1061, s2053 and s8305 from the High End Computing (HEC) division of NASA. We thankfully acknowledge the support of the Research Council of Norway through grant 230938/F50 and through grants of computing time from the Programme for Supercomputing. 

\bibliographystyle{aa}
\bibliography{aamnemonic,collectionbib}

\end{document}